\documentclass[preprint]{aastex}
\slugcomment{submitted to {\it The Astronomical Journal}}
\begin{document}
\title{$vbyCa$H$\beta$ CCD Photometry of Clusters. VIII. The Super-Metal-Rich, Old Open Cluster NGC 6791}
\author{Barbara J. Anthony-Twarog, Bruce A. Twarog, and Lindsay Mayer}
\affil{Department of Physics and Astronomy, University of Kansas, Lawrence, KS 66045-7582}
\affil{Electronic mail: bjat@ku.edu, btwarog@ku.edu, mayer56@ku.edu}

\begin{abstract}
CCD photometry on the intermediate-band $vbyCa$H$\beta$ system is presented for the metal-rich, old open cluster, NGC 6791. Preliminary analysis led to [Fe/H] above +0.4 with an anomalously high reddening and an age below 5 Gyr. A revised calibration between $(b-y)_0$ and [Fe/H] at a given temperature shows that the traditional color-metallicity relations underestimate the color of the turnoff stars at high metallicity. With the revised relation, the metallicity from $hk$ and the reddening for NGC 6791 become [Fe/H] = +0.45 $\pm$ 0.04 and E$(b-y)$ = 0.113 $\pm$ 0.012 or E$(B-V)$ = 0.155 $\pm$ 0.016. Using the same technique, reanalysis of the photometry for NGC 6253 produces [Fe/H] = +0.58 $\pm$  0.04 and E$(b-y)$ = 0.120 $\pm$ 0.018 or E$(B-V)$ = 0.160 $\pm$ 0.025. The errors quoted include both the internal and external errors. For NGC 6791, the metallicity from $m_1$ is a factor of two below that from $hk$, a result that may be coupled to the consistently low metal abundance from DDO photometry of the cluster and the C-deficiency found from high dispersion spectroscopy. E$(B-V)$ is the same value predicted from Galactic reddening maps.  With E$(B-V)$ = 0.15 and [Fe/H] = +0.45, the available isochrones predict an age of 7.0 $\pm$ 1.0 Gyr and an apparent modulus of $(m-M)$ = 13.60 $\pm$ 0.15, with the dominant source of the uncertainty arising from inconsistencies among the isochrones. The reanalysis of NGC 6253 with the revised lower reddening confirms that on both the $hk$ and $m_1$ metallicity scales, NGC 6253, while less than half the age of NGC 6791, remains at least as metal-rich as NGC 6791, if not richer.

\end{abstract}

\section{INTRODUCTION}
This is the eighth paper in an extended series detailing the derivation of fundamental parameters in star clusters using precise intermediate-band photometry to identify probable cluster members and to calculate the cluster's reddening, metallicity, distance and age. The initial motivation for this study was provided by \citet{tat97}, who used a homogeneous open cluster sample to identify structure within the galactic abundance gradient. The evolution of the argument surrounding the exact nature of that structure has been laid out in the previous papers in this series and the reader is referred to the references cited in the most recent papers \citep{at04,at05,at06,tw06} for more details. In its basic form, the question at hand is the reality of a single slope defining the decline in [Fe/H] with galactocentric distance from the Galactic bulge to a location more that 20 kpc from the Galactic center, as opposed to at least two zones with shallower slopes coupled by a distinct transition zone located 1 to 3 kpc beyond the solar circle. The suggestion of another transition zone interior to the solar circle has recently been made by \citet{lu06}.

Detailed justifications of the adopted observational approach have been given in early papers in the series \citep{at00a,at00b,tw03} and will not be repeated. However, as with any photometric or spectroscopic technique, it is critical that the procedures and calibrations be constantly tested and reevaluated as the sample expands and the observations extend to parametric ranges beyond those originally defined by the technique. A relevant example is provided by the revised calibration by \citet{fj02} of the moderate-resolution spectroscopic technique used originally to study the old cluster sample in \citet{fj}. 

The focus of this paper is the old, metal-rich open cluster, NGC 6791. Interest in and analysis of this object have grown dramatically with time due to the repeated claims that it falls near the upper limits in the range of age and metallicity among open clusters, two crucial parameters of importance to those studying both Galactic and stellar evolution. Its role within any coherent model of the chemical evolution of the disk could be downplayed, if not ignored, if it remained the sole example of an old cluster with exceptionally high metallicity though, again, the question of its exact metallicity has itself been a matter of controversy. This situation changed with the discovery of a second, old open cluster interior to the sun, NGC 6253, with indications of a metallicity at least twice that of the sun \citep{br97,sa01}. This claim was confirmed and quantified within this photometric program using CCD intermediate-band photometry \citep{tw03}, implying a metallicity at least as high as that claimed for NGC 6791, i.e., [Fe/H] above +0.4. Exactly how high the metallicity went was difficult to tell because the intermediate-band system that generated the reddening and the estimated abundance had never been adequately tested/calibrated with field stars for the range of indices found at the turnoff of NGC 6253. It was hoped, therefore, that the observations of NGC 6791 would shed light not only on the basic properties of the cluster, but also upon the viability of the photometric results for NGC 6253.

First analysis of the intermediate-band CCD data for NGC 6791 in 2004 clearly indicated a cluster [Fe/H] of +0.4 or higher, as reported by \citet{act04}. However, the complementary result was a reddening value of E(B-V) larger than predicted by other techniques by $\sim$ 0.10 mag, requiring an anomalously low age (4 to 6 Gyrs) and a CMD morphology that could not be matched by isochrones employing scaled-solar abundances. The fact that a virtually identical problem arose with the CMD morphology of NGC 6253 led us to the conclusion that the issue is somehow tied to the unusually high metallicity of the two clusters, rather than an error in the photometry. As we will demonstrate below, a revision of the intermediate-band calibrations for high metallicity stars from an expanded sample of field stars resolves the discrepancies for both clusters.   

Section 2 contains the details of the $vbyCa$H$\beta$ CCD observations, their reduction and transformation to the standard system. In Sec. 3 we discuss the CMD and begin the process of identifying the sample of probable cluster members. Sec. 4 contains the derivation of the fundamental cluster parameters of reddening and metallicity. Sec. 5 details and resolves a problem with the derivation of the reddening using traditional color-temperature relations.   Sec. 6 discusses the the revised cluster parameters, the age and distances of NGC 6791 and NGC 6253, and their potential relevance for our understanding of Galactic evolution. Sec. 7 summarizes our conclusions.

\section{THE DATA}

\subsection{Observations: CCD $vbyCa$H$\beta$}
New photometric data for NGC 6791 were obtained using the 0.9-meter WIYN telescope on Kitt Peak. The S2KB $2048 \times 2048$ CCD mounted at the the $f/7.5$ focus of the telescope provides a field approximately $20'$ on a side. For five photometric nights between 2004 May 27 and May 31, NGC 6791 was the sole program target for frames taken with our own $3\times 3$ inch H$\beta$ and $Ca$ filters and $4 \times 4$ inch $y,b,$ and $v$ filters borrowed from from KPNO.  

Processing of all frames through bias subtraction, trimming and flat fielding was accomplished at the University of Kansas using standard IRAF routines.  Dome flats were used for all filters.  The analysis of photometry for the cluster is based on 13 to 15 frames in each of the six filters, with total exposure times of 47, 64, 127, 168, 60 and 160 minutes for the $y, b, v, Ca$, H$\beta$ wide and narrow filters, respectively.

\subsection{Reduction and Transformation}
Previous papers in this series describe the procedures used to produce instrumental photometry with high internal precision using IRAF Allstar routines, with \citet{at00a} providing the most thorough description of our approach.  The accuracy with which our instrumental photometric indices are transformed to the standard systems is a particularly important concern for NGC 6791.  These procedures, too, have been extensively described in previous papers, especially \citet{at04}. 

Our calibrations are based on aperture photometry of stars in the program cluster and of standard stars in the field and in available open clusters.  For all of our May 2004 nights, the seeing was 1.5 to 2.0\arcsec, permitting the use of apertures 6\arcsec\  in radius surrounded by a sky annulus of comparable area. A number of sources were consulted for standard values, including the catalog of \citet{tat95} for $V$, $b-y$ and $hk = (Ca-b) - (b-y)$ indices, catalogs of $uvby$H$\beta$ observations by \citet{ols83,ols93,ols94}, and compilations of H$\beta$ indices by \citet{hm98}, \citet{sn89}, and \citet{cb}.  Stars in NGC 6633 were also observed, using photometry by \citet{ed76}.

Following standard procedures for Str\"omgren photometry, separate $b-y$ calibrations were derived for cooler dwarfs, with warmer dwarfs and giant stars comprising a separate calibration class; calibration equations for $m_1 = (v-b) - (b-y)$ are even further subdivided into calibration class (cooler dwarfs, warmer dwarfs and giants). Additional details about the calibration equations are included in Table 1.  After an examination of observed and standard indices for all of the data from all photometric nights, a common slope was determined for each calibration equation, leaving the zero point to be determined for each calibration equation from each nights' standard star observations.  For most calibration equations, the zero point was determined based on standard values from the field star catalogs, the exception being the $m_1$ calibration equations for which standards in NGC 6633 were used as well.  

The extension of calibration equations to the merged profile-fit photometry is facilitated by determining the average differences between the profile-fit indices for stars in NGC 6791 and indices determined from aperture photometry on each photometric night.  It was possible to determine the average difference between the  aperture and profile-fit indices for 50 to 90 stars in NGC 6791 with a dispersion of 0.02 mag or less, so that the ``aperture corrections" for each index could be determined to very high precision.  

Photometric errors as a function of $V$ for the various indices are illustrated in Figs. 1 and 2. Final photometry for 6313 stars with at least two observations in $y$ and $b$ is given in Table 2. Identifications and XY coordinates are on the system found in the WEBDA Cluster Data Base. For those stars not included in the WEBDA database, an identification number has been created beginning with 10000.

\subsection{Comparison to Previous Photometry: Broad-Band $V$}
While there have been a number of broad-band CCD studies of the cluster, we will focus on a comparison of our $V$ photometry with that of \citet{stet}, where a comprehensive comparison is made with all the previous surveys of the cluster. As in the previous papers in the series, we will compare the residuals in the photometry for the entire sample and for the brighter stars alone since the reliability of the latter sample is more relevant to the determination of the cluster parameters.  Data for \citet{stet} have been taken from the WEBDA Cluster Data Base. 

For 3980 stars in common with \citet{stet}, the average residual in the sense (Table 2 - STE) is $-0.009 \pm 0.043$, excluding only stars with residuals greater than 0.20 mag. If we make a more restrictive cut to the residuals of  0.10 mag, the average from 3844 stars is only slightly changed, $-0.007 \pm 0.036$. Moving to the brighter stars of relevance for the discussion of the cluster parameters, for 2262 stars with $V <$ 18.3, the residuals average $-0.018 \pm 0.132$, with no exclusions. If we exclude stars that deviate from the mean by more than 0.075 mag, the average residual for 2112  stars becomes $-0.003 \pm 0.027$. No color term of significance, based upon the low correlation coefficients for the attempted fits, is found among the residuals in any comparison.  The fraction of stars that are excluded by the cut among the brighter sample isn't negligible, but is readily explained upon closer examination. Of the 150 stars dropped due to the cut, only 10 have residuals of +0.076 or greater; the remainder are all negative. The majority of the deviants are stars with $V >$ 18.0 in WEBDA. This asymmetry is what one would expect if two stars with comparable positions, one faint and one bright, are both identified as the same brighter star. Thus, the positional match between our CCD data and that of the WEBDA data base is comparing the brighter companion in our sample with both stars in the WEBDA catalog, even if the fainter star isn't present in our survey. We conclude that our $V$ magnitudes and those of \citet{stet} are on the same system to better than 0.01 mag.

\section{THE COLOR-MAGNITUDE DIAGRAM}

\subsection{Thinning the Herd: Proper-Motion Members}
With a cluster like NGC 6791, there are two competing factors that impact the selection of probable cluster members. On one hand, the cluster is unusually rich for an open cluster, particularly for a very old one. In the absence of a CMD, it could easily be misclassified as a low density globular cluster. On the other hand, the cluster is located interior to the solar circle, somewhat toward the galactic center ($l, b =$ 70\arcdeg, +11\arcdeg), superposing it upon an annoyingly rich background of field stars. In past papers in the series, the initial step in maximizing the member sample has been to restrict the sample to stars with high precision photometry within the cluster core. 

The CMD for all stars with two or more observations in both $b$ and $y$ is shown in Fig. 3. Stars with errors greater than $\pm$0.010 in $b-y$ are presented as crosses. The richness of the cluster is readily apparent as is the extensive field star contamination. The giant branch and clump are well-defined, but the probable location of the turnoff is only discernible through the identification of the base of the giant branch. The field star F-dwarf population forms a continuous band between $b-y$ = 0.4 and 0.6, growing more populous at fainter magnitudes and making delineation of the true cluster turnoff and main sequence almost impossible. Additionally, although the photometry extends 2.5 mag fainter than the likely cluster turnoff, the typical errors in $b-y$ rise above 0.010 mag near $V \sim 17.3$, the expected location of the cluster turnoff. Thus, unlike previous open clusters in the series, we will not have an extended range of turnoff stars with photometric errors in the individual indices well below 0.01 mag. The significant age of the cluster is obvious from the length of the giant branch while high metallicity is suggested by the observed slopes of the upper giant branch and the subgiant branch. 

In an attempt to reduce field star contamination, we next restrict the data to the spatial core of the cluster. We applied the same technique based upon radial star counts used in past papers in this series to define the probable center of the cluster. Stars with $V$ brighter than 19.0 within 300 WEBDA coordinate units (corresponding to 3.85 \arcmin) of the newly defined coordinate center were retained. The improvement in the homogeneity of the sample is illustrated by Fig. 4, where the core stars are plotted using the same symbols as in Fig. 3. While the delineation of the cluster CMD is greatly enhanced, the rich population of stars above the turnoff from the blue straggler region to the giant branch is an indication that serious contamination of the cluster sample by field stars, especially near the turnoff, remains a legitimate concern.

To eliminate the probable cluster non-members, we turn to the proper-motion analysis by Cudworth (private communication). Preliminary results from the initial analysis of the field of NGC 6791 were presented by \citet{cat}, but the measured proper motions and the separation of the field from the cluster have been improved since that first discussion and Dr. Cudworth has kindly made his current results available to us. Since we are interested in obtaining the purest sample of cluster stars for our derivation of the cluster parameters, we place the relatively tight restriction of  a membership probability of 90$\%$ or above. We gain, however, by including stars that meet this criteria anywhere within the field of the CCD frame, not just the cluster core. The resulting CMD is shown in Fig. 5 where the improvement over Fig. 4 is dramatic. It is clear that there is a rich population of blue stragglers/binaries just above the cluster turnoff, but the red giant branch is narrow and well-delineated from the base to almost 0.5 mag above the clump. The number of members identified on the main sequence declines steadily below $V \sim 18.4$, undoubtedly due to increasing uncertainty in the membership probability determination coupled with the restrictive cut adopted here.

\subsection{Thinning the Herd: CMD Deviants}

For purposes of optimizing the derivation of the cluster reddening and metallicity, our interest lies in using only single stars that evolve along a traditional evolutionary track and that have indices in a color range where the intrinsic photometric relations are well defined. With this in mind, we next cut the sample to include only stars in the magnitude range from $V$ = 17.3 to 18.3 and the color range from $b-y$ = 0.5 to 0.65, thereby eliminating blue stragglers, some binaries, evolved subgiants, the red giant branch, and fainter stars with larger probable photometric errors. The 774 stars of Fig. 5 are thereby reduced to 376. We make a further restriction that the photometric errors be less than or equal to 0.020, 0.030, 0.040, and 0.030 for $b-y$, $m_1$, $hk$, and H$\beta$, respectively. This reduces the sample to 363 stars.

Given the proper-motion data, it is highly probable that the majority of stars at the cluster turnoff are members, though not necessarily single stars. As in past papers in the series, it is valuable to identify and remove the likely binary systems from the sample to avoid distortions of the indices through either anomalous evolution or simple photometric combinations of the light for stars in significantly different evolutionary states. For stars in the vertical portion of the turnoff, separation of the stars into two parallel sequences, one single and one composed of binaries, becomes a challenge because the single stars evolving away from the main sequence produce an evolutionary track that curves toward and crosses the rich binary sequence composed of unevolved pairs. 

To enhance our ability to delineate the single and binary sequences, we use $v-y$, a color index with a larger baseline in wavelength and greater sensitivity to temperature change than $b-y$. Moreover, $v$ is dominated by metallicity effects rather than surface gravity and all the stars within the cluster supposedly have the same [Fe/H].  For the 363 stars remaining after our previous cuts, the $V, v-y$ diagram is illustrated in Fig. 6.  Despite the almost vertical nature of the turnoff, it is apparent that there is an asymmetric distribution of stars weighted to the redder side of the diagram. The points below the subgiant branch are likely contributed by binaries shifted vertically by the combined color and luminosity of the pair, in some cases produced by stars that are not represented by any single-star analogs because they fall fainter than the $V$ limit of the sample. Keeping in mind that the membership probability cut will still lead to the inclusion of $\sim$5$\%$ non-members, some stars, e.g. those that lie blueward of the main band, can be either non-members or stars with anomalously poor photometry. Finally, the extended color range of the subgiant branch highlights the stars whose colors are most affected by evolution away from the turnoff. Those stars that deviate the most from the single-star sequence are tagged as filled circles and will be excluded in further discussions. 

As we have done in the past, a check on the classification has been made using the $V,(Ca-y)$ diagram, with excellent correspondence with the structure seen in Fig. 6. The fact that there is an almost one-to-one correspondence between the reddest and bluest stars in both diagrams gives us confidence that we are measuring true color differences rather than random photometric scatter since the the $Ca$ and the $v$ filters are reduced and calibrated independently of each other. Though there may be a handful of stars that could be added to the list of deviants based upon their position in the $V- (Ca-y)$ diagram, we will exclude only those 44 stars already tagged in Fig. 6.

\section{FUNDAMENTAL PROPERTIES: REDDENING  AND METALLICITY}

With 319 stars selected in a tightly constrained region of the cluster turnoff, the normal procedure would be to calculate individual reddening values, correct for a cluster mean reddening, and use the individual metallicity estimates to define a cluster mean [Fe/H]. However, unlike past analyses in this series, we have retained photometry with larger than average errors in order to reach one magnitude below the brightest stars at the turnoff, expanding the sample size by a factor of three compared to previous work. Exploiting the larger sample size, we have averaged indices for stars sorted in 0.1 mag bins in $V$ between 17.3 and 18.3. The averages were derived using weights based upon the inverse square of the standard error of the mean for the individual indices. The resulting averages and their uncertainties are listed in Table 3. It should be emphasized that the table values would have been virtually identical had we used unweighted averages.

Additionally, we will derive the fundamental parameters in two phases, the first based upon the calibrations used in past papers in this series. The anomalous implications of the results will point to the need for a new calibration of the intermediate-band system at high metallicity, leading to a revised set of cluster parameters in the second phase.

\subsection{Reddening: Phase 1}
Derivation of the reddening from intermediate-band and H$\beta$ photometry assumes that H$\beta$ is reddening and metallicity independent, while metallicity and reddening affect the remaining intrinsic colors. Reddening is derived for a range of assumed values for the metallicity index $\delta$$m_1$($\beta$), then the metallicity index is derived for a range of assumed reddening values. Only one combination of $E(b-y)$ and $\delta$$m_1$ will simultaneously satisfy both relations. The primary decision is the choice of the standard relation for H$\beta$ versus $b-y$ and the adjustments required to correct for metallicity and evolutionary state. The two most commonly used relations are those of \citet{ols88} and \citet{ni88}. As found in previous papers, both produce very similar if not identical results. 

A modest twist in the current investigation comes from the lack of $c_1$ due to the absence of the $u$ filter. For the intrinsic colors, most relations include a term weakly dependent upon $\delta$$c_1$($\beta$), the difference between the observed index at a given value of H$\beta$ and the standard relation for unevolved stars; the more evolved a star is, the larger the value of $\delta$$c_1$($\beta$). Fortunately, because we have removed the most probable sources of confusion between evolution and binarity in the CMD, any deviation from the unevolved main sequence in the CMD for mean points in Table 3  must be attributed to evolution. As a crude approximation to provide some insight into the effect of $c_1$, we have calculated the position of each mean point in the CMD above the zero-age main sequence at the same color, $\delta$$M_V$, using the broadband photometry of \citet{stet} for the cooler stars to define the unevolved main sequence and fix the location in $V$ of the unevolved main sequence at the color of the turnoff. The value of $\delta$$M_V$($\beta$) has then been converted to $\delta$$c_1$($\beta$) and thus to $c_1$.  We emphasize that the inclusion of these terms has a minor impact on both the reddening and metallicity estimates, so the added scatter created by the possible errors in the $c_1$ estimates is modest.

Processing the binned data of Table 3 through both relations generates $E(b-y)$ = 0.160 $\pm$0.010 (s.d.) with \citet{ols88} and $E(b-y)$ = 0.164 $\pm$0.009 (s.d.) with \citet{ni88}, with $\delta$$m_1$($\beta$) = --0.026 $\pm$0.003 (s.e.m.). As a compromise, we will take the simple average of the two and use $E(b-y)$ = 0.162 $\pm$0.003 (s.e.m.) or $E(B-V)$ = 0.225 $\pm$0.004 (s.e.m.) in the analyses that follow.

\subsection{Metallicity from $m_1$ and $hk$} 
Given the reddening, the derivation of [Fe/H] from the $m_1$ index is as follows. The $m_1$ index for a star is compared to the standard relation at the same color/temperature and the difference between them, adjusted for possible evolutionary effects, is a measure of the relative metallicity, i.e., $\delta$$m_1$ = $m_{1STAN}$ - $m_{1OBS}$ . Use of H$\beta$ rather than $b-y$ to define the standard relation decouples the impact of errors in the indices ($m_1$ includes the $b$ and $y$ filters) while making use of the lack of metallicity and reddening sensitivity for H$\beta$. 

After correcting each mean index for the effect of $E(b-y)$ = 0.162 and deriving the differential in $m_1$ relative to the standard relation at the observed H$\beta$,  the average $\delta$$m_1$ for 10 bins is $-0.026$ $\pm$0.002 (s.e.m.), which translates into [Fe/H] = +0.37 $\pm$0.029 (s.e.m.) for the calibration as defined in \citet{ni88} and adopted in previous papers. The zero-point of the H$\beta$ metallicity calibration has been fixed to match the adopted value for the Hyades of [Fe/H] = +0.12.

The primary weakness of metallicity determination with intermediate-band filters is the sensitivity of [Fe/H] to small changes in $m_1$; the typical slope of the [Fe/H]/$\delta$$m_1$ relation is 12.5. Even with highly reliable photometry, e.g., $m_1$ accurate to $\pm$0.015 for an individual faint star, the uncertainty in [Fe/H] for an individual star is $\pm$0.19 dex from the scatter in $m_1$ alone. When potential photometric scatter in H$\beta$ and $b-y$ are included, errors at the level of $\pm$0.25 dex are common, becoming even larger for polynomial functions of the type adopted by \citet{no04}. The success of the adopted technique depends upon both high internal accuracy and a large enough sample to bring the standard error of the mean for a cluster down to statistically useful levels, i.e., below $\pm$0.10 dex. Likewise, because of the size of the sample, we can also minimize the impact of individual points such as binaries and/or the remaining nonmembers, though they will clearly add to the dispersion.

The alternative avenue for metallicity estimation is through the $hk$ index. The $hk$ index is based upon the addition of the $Ca$ filter to the traditional Str\"{o}mgren filter set, where the $Ca$ filter is designed to measure the bandpass that includes the H and K lines of ionized calcium. The design and development of the $Caby$ system have been laid out in a series of papers that have been cited extensively in past papers in the series. Though the system was optimally designed to work on metal-poor stars and most of its applications have focused on these stars \citep{atc95,bd96,at00}, of particular importance for NGC 6791, early indications that the system retained its metallicity sensitivity for metal-rich F dwarfs have been confirmed by observation of the Hyades and analysis of nearby field stars \citep{at02}. What makes the $hk$ index so useful for dwarfs, even at the metal-rich end of the scale, is that it has half the sensitivity of $m_1$ to reddening and approximately twice the sensitivity to metallicity changes. The metallicity calibration for F stars has been done by defining a $\delta$$hk$ = $hk_{STAN}$ - $hk_{OBS}$ for stars relative to a standard relation using both $b-y$ \citep{at02} and H$\beta$ \citep{tw03} as the temperature index, analogous to the $\delta$$m_1$ parameter for $uvby$ photometry. Linear relations between $\delta$$hk$ and [Fe/H] are then generated for each temperature range. Only minor revisions have been made to the original calibrations as the field star abundance data have improved and expanded.  For the reasons noted above, we will again focus on the metallicity index defined relative to H$\beta$.

Using the binned data of Table 3 and E$(b-y)$ = 0.162, the metallicity for NGC 6791 is +0.50 $\pm$ 0.025 (s.e.m.). The value is higher than derived from $m_1$, but a difference of this size is reasonable given the higher sensitivity of $m_1$ to reddening and potential errors in the photometric zero-point; an increase of 0.01 mag in the $m_1$ scale would bring the two results into perfect alignment. A weighted average of the two abundance estimates produces [Fe/H] = +0.44 for NGC 6791.

\section{THE PROBLEM: AGE AND REDDENING COMPARISONS}

At this point, there seems to be little doubt that the cluster is very metal-rich. The logical next step would be an age determination based upon a CMD fit to an appropriate set of isochrones. The problem that immediately emerges is identical to what was encountered in the analysis of NGC 6253 \citep{tw03}. The broad-band CMD of NGC 6791, when corrected for a reddening of E$(B-V)$ = 0.225, is morphologically incompatible with all isochrones with scaled-solar abundances and [Fe/H] of +0.40. One of several issues is that the turnoff color requires an age for NGC 6791 below 4 Gyr, inconsistent with the lack of a hydrogen-exhaustion-phase hook; another is that the observed giant branch is too blue by 0.15 mag relative to the isochrone. (For the analogous problem with NGC 6253, see Fig. 9 of \citet{tw03}.) An {\it ad hoc} solution, following the same pattern laid out for NGC 6253, is to assume that the [Fe/H] value for NGC 6791 is actually lower than derived from the photometry, but distorted by an abundance distribution that is not scaled-solar, specifically one that has alpha-element enhancements. Fig. 7 shows such a match using the isochrones of \citet{pad} incorporating modest levels of overshoot superposed on the cluster photometry of \citet{stet}. Because of the prescription used to derive the overshoot parameters, it isn't possible to quantify the degree of overshoot using a single number such as the fraction of a pressure scale height.  The $(B-V)$ photometry has been reddening-corrected and the individual isochrones have been adjusted with distance moduli designed to bracket the unevolved main sequence. The interpretation would be that by interpolating between the two selected sets, the approximate properties for NGC 6791 should be an age near 8 Gyr with [Fe/H] $\sim$ +0.2 but alpha-enhanced abundances and an apparent distance modulus of $(m-M)$ $\sim$ 13.6.

\subsection{Comparison to Previous Determinations: Reddening and Metallicity}
A non-solar elemental distribution for NGC 6791 was invoked to reconcile the estimated reddening of E$(B-V)$ above 0.20 with a cluster that is significantly old (7 Gyrs or more). The morphology of the CMD, the relative position of the turnoff, giant branch, and red giant clump, and the luminosity function all require an age considerably greater than a few billion years. Exactly how high the age must be has been one of the unresolved ongoing controversies regarding the cluster and a driving force behind the need for improved cluster parameters of reddening and metallicity. Among the more recent derivations of key cluster parameters, the reader is referred to \citet{stet,ca05,ca06,or06,gr06} and the many citations therein addressing these longstanding questions. 

The difficulty in deriving these parameters is that most techniques used to measure them are dominated by their interdependence. The intrinsic colors for a given class of stars will depend upon their assumed age and metallicity, but the age can't be reliably known without the metallicity which, in turn, requires knowledge of the reddening. These concerns extend back to the earliest attempts by \citet{ki} to define the cluster reddening at E$(B-V)$ = 0.22 using spectral classification and the observed colors of 16 cluster stars, under the assumption that the cluster stars had the same approximately solar metallicity as the stars that defined the standard color relation. This assumption was repeated or derived for a variety of attempts over the next 20 years, resulting in estimates ranging from E$(B-V)$ = 0.10 from DDO photometry \citep{ja84} to 0.13 \citep{hc81} and 0.20 \citep{at85} from photoelectric $UBV$ data. Although \citet{mj94} adopted the more traditional approach of using $UBV$ color-color diagrams to obtain E$(B-V)$ = 0.10, over the last 20 years examination of the cluster parameters has been dominated by comparisons to theoretical isochrones and isochrone morphology, either assuming a value for the reddening or adjusting the reddening and metallicity to optimize the match to various CMD and color-color diagrams \citep{ka90,de92,me93,ca94,ga94,tr95,kr95,ch99,de04,stet,ca05,ca06}. Within this list of citations, prior to the investigation by \citet{ch99}, the adopted and/or derived reddening estimates ranged between E$(B-V)$ = 0.15 and 0.24, with an average value of 0.20, while the assumed metallicity went between [Fe/H] = 0.00 and 0.44. Since the work of \citet{ch99}, estimates for the reddening have ranged between E$(B-V)$ = 0.09 and 0.17, with [Fe/H]  between 0.3 and 0.5. The one constant in all this has been that morphologically-based CMD parameters have invariably implied that NGC 6791 is older than NGC 188, though the size of the differential and the absolute scale have varied with the choice of stellar models \citep{ja83,at85,ta89,jp94,cc94,sa04}.

The importance of this modest history of our supposed understanding of the cluster reddening is that if the cluster is truly metal-rich, i.e., [Fe/H] above 0.3, the growing sample of broad-band photometry ranging from the optical to the infrared can only be made consistent with a large age for the cluster if E$(B-V)$ lies between 0.06 and 0.18, bounded on the lower side by E$(B-V) = 0.09 \pm 0.03$ from \citet{stet} and the maximum value consistent with E$(B-V) = 0.14 \pm 0.04$ from \citet{ca05}. One independent piece of evidence supporting this range is supplied by the reddening maps of \citet{sc98}, where the Galactic reddening in the direction of NGC 6791 is E$(B-V)$ = 0.154, a value that should represent an upper limit to the cluster measure.

Can we assume that [Fe/H] is above +0.3? Recent high dispersion spectroscopic work consistently generates super-metal-rich values for NGC 6791, including [Fe/H] = +0.35 \citep{or06}, +0.39 \citep{ca06}, and +0.47 \citep{gr06}. It is important to remember, however, as is often the case in spectroscopic studies of a small number of objects, that it can be a challenge to determine if different investigations are on the same metallicity scale, independent of the details of the temperature scale and spectroscopic analysis. However, one can make use of the intermediate-band photometry, specifically $hk$. One of the advantages of the $hk$ index that has been repeatedly emphasized is its weak dependence on errors in reddening. In the previous section, [Fe/H] = +0.50 was derived assuming E$(b-y)$ = 0.162 or E$(B-V)$ = 0.225, well beyond the upper limit discussed above. If we assume that the reddening must lie between E$(B-V) = 0.16$ and 0.06 (E$(b-y)$ $\sim$ 0.12 to 0.04), the [Fe/H] range from the $hk$ index is +0.48 to +0.43, i.e., NGC 6791 is definitively super-metal-rich.

\subsection{Solutions: The Reddening - Phase 2}
Since our abundance estimates, even those derived from $hk$, are affected by an H$\beta$ scale that appears to imply
reddening estimates for this cluster that are too large by 0.06 to 0.16 mag in E$(B-V)$, the reader might be justified in questioning the validity of the photometry itself.  We carefully considered but rejected the explanation that these problems are caused by significant errors in the zeropoints of the photometric calibrations.  Independent observations might prove otherwise, but the size of the errors required to cause such a discordant effect is two to three times larger than our most generous assessments of the photometric errors in the zero-points. 

If the photometry is reliable and the stars within NGC 6791 are not peculiar in the sense of having non-scaled-solar elemental abundance distributions, other options to explain the high reddening value had to be considered. As discussed in \citet{tw03}, it is possible that at super-metal-rich abundance levels, the standard relations linking H$\beta$, $b-y$, and $m_1$ do not adequately predict the intrinsic $(b-y)_0$ colors of stars at all temperatures.  Given that the functions we have regularly used \citep{ols88, ni88} were derived at a time when the number of stars with [Fe/H] significantly above +0.2 was small and these stars were scattered over a wide range in color, it is not impossible that for [Fe/H] above 0.4, deviations from the true colors would manifest themselves.

Fortunately, in conjunction with the analysis of the data for NGC 6791, a complementary project was under way to redefine the metallicity calibration for the $hk$ system. With the publication of the homogeneous, high-dispersion, spectroscopic abundance catalog of \citet{vf} (hereinafter referred to as VF), the program was adapted to redefine the metallicity calibration for the traditional $uvby$ system for cooler F and G dwarfs in order to ensure that the $hk$ and $m_1$ systems defined an internally consistent abundance scale. In the course of the latter effort, the solution to the reddening problem in NGC 6791 (and ultimately NGC 6253) became apparent. The extensive details on the recalibration of the extended Str\"{o}mgren system will be presented in a future paper in this series. For now, we will supply only the materials relevant to the question at hand, the reddening and metallicity of NGC 6791.

The success of any statistical analysis of a large data sample generated by the merger of data from multiple sources, particularly one tied to the definition of a metallicity scale, is invariably driven by the degree of homogeneity that can be imposed upon the merged components. Approaches to such mergers for abundance catalogs are varied, ranging from the very basic \citep{ca01} to the more elaborate \citep{bt}. Our approach lies somewhere in the middle and represents a modified version of the technique used in the recalibration of the DDO system \citep{tw96}. The critical component in our approach is the existence of a large base catalog of homogeneous high dispersion spectroscopic abundances obtained by one group using the same reduction technique for all stars. For the red giant discussion, the data of \citet{mw90} proved ideal; for the dwarfs, the exquisite sample of VF has proven invaluable. Given the base catalog, the next step is the transformation of additional sources to the base standard. For the giants, issues of reddening were important due to the range in distance among the sample, the impact upon the temperature scale, and ultimately the abundance. For the dwarfs, virtually all of the stars are within 100 pc and, on average, reddening effects should be small to nonexistent. For the sake of simplicity, we assume that whatever corrections have been made for reddening within the individual sources, even if they contradict each other, are correct and merely add to the scatter among the comparisons. In contrast with the giants, however, with only a few exceptions, all the stars in the dwarf catalog have been observed by Hipparcos and therefore have some form of distance and absolute magnitude estimate. These have been used in some cases to generate surface gravity estimates used in analysis of the spectra, while others have used the internal consistency of the line analysis to define a log g. In deriving our transformations to the base catalog, we have therefore attempted to optimize the match by using the following relation:

\centerline{$[Fe/H]_{FV} = a\ [Fe/H]_{ref} + b\ log\ T_e + c\ log\ g + d$}

In many cases, the significance of the surface gravity and effective temperature terms was negligible and they were dropped from the final transformations. It's encouraging that the linear slopes were normally close to 1.0 and the offsets consistently below 0.10 dex.

To create the catalog, the literature in recent years was surveyed and papers with high-dispersion spectroscopic abundances for large samples of dwarfs were cross-correlated with VF.  Additionally, we included the data from the first large spectroscopic analysis of the chemical evolution of solar neighborhood by \citet{ed93}. If the source catalog exhibited adequate overlap with VF and, after applying a transformation equation of the form noted above, produced a scatter among the residuals between the source catalog and VF below 0.06 dex, the catalog was retained. Adopting an approximate estimate of $\pm$0.025 dex as the typical uncertainty within the catalog of VF, the error in the residuals was used to define an approximate internal error for each catalog. The data for each catalog were then transformed to the system of VF and the abundances for each star averaged using a weighting scheme based upon the inverse of the adopted internal errors for each catalog. 

In addition to VF and \citet{ed93}, the catalogs included to date in the merged sample are \citet{ap04,ch06,fu98,gi06,hu05,la03,lh06,re03,re06,tak5}. It is expected that more surveys will be included in the final sample used to redefine the photometric calibrations but, for now, the merged catalog contains abundances for 1574 stars, 1039 of which were found in the original sample of VF.

The large sample of supposedly homogeneous abundances was then cross-matched with the published catalogs of $uvby$H$\beta$ data. The only $uvby$ indices used were those from the catalogs of \citet{ols83,ols93,ols94}, complemented by the data from the same program as presented by \citet{no04}. Likewise, the primary source of H$\beta$ photometry has been the summary of data on a common system as presented by \citet{no04}, but expanded to include data for some stars from the catalog of \citet{hm98}. The inclusion of the last source was done to generate as large an overlap as possible for the analysis at hand. $uvby$ indices are readily available for stars across a wide range in temperature, metallicity, and surface gravity. H$\beta$ photometry is less commonly obtained and is often ignored for stars assumed to be too cool to supply any useful information via the index. Since we are dealing with stars that appear red in $b-y$ because they are metal-rich rather than cool, it is important to have as many field stars as possible that meet the same criteria, i.e., appear to be mid to late G-type stars from their $b-y$ colors, but are hotter than the sun as defined by H$\beta$. Of the 1574 stars in the abundance catalog, 1088 have $uvby$ indices from the sources noted above. Of these, H$\beta$ photometry is available for 878 stars.

Returning to NGC 6791, the data of Table 3 cover the almost vertical turnoff region of the CMD, resulting in a very modest range in H$\beta$ among the stars, a combination of the slight curvature of the CMD and the internal photometric scatter.  The weighted average of the H$\beta$ values at the turnoff is 2.597 $\pm$ 0.007; if the brightest point of the turnoff is excluded, the average changes to 2.598 $\pm$ 0.005. For comparison purposes with our catalog, we have selected all stars with H$\beta$ between 2.596 and 2.600, inclusive. In Fig. 8 we plot the trend of $b-y$ with [Fe/H] for the 54 stars that meet our criteria.  The trend of increasing $b-y$ color with increasing [Fe/H] is obvious, though the linear trend appears to bottom out for [Fe/H] below --0.35, i.e., stars with [Fe/H] below --0.35 have effectively the same color. The number of stars at the low metallicity end is small due to the restriction imposed by the narrow range in H$\beta$. If we expanded the range to increase the sample, the very shallow slope below [Fe/H] = --0.35 is confirmed, but the range among metal-rich stars expands due to the changing color-[Fe/H] relation with temperature. For stars with [Fe/H] above --0.35, the observed scatter has a variety of plausible sources, including photometric scatter in $b-y$ or H$\beta$, uncertainties in [Fe/H], though the typical star should have an error below 0.035 dex, and reddening, which has been neglected. The last factor will shift a star vertically upward in the plot. Finally, as is the case at the turnoff of NGC 6791, the stars may be in a range of evolutionary state or surface gravity. The one star that deviates by a significant amount from the general trend ($b-y$ $\sim$ 0.44) is HD 14412. There are multiple observations of $b-y$ for this star, all of which agree well that it is a mid-G dwarf. Its parallax indicates a distance within 20 pc of the sun, so no correction for reddening should be required. There are two published H$\beta$ observations from the same investigators that differ by 0.017 mag. The four transformed spectroscopic abundances cover a range of only 0.09 dex. The problem appears to be that the H$\beta$ observations of this star imply too high a temperature for the $b-y$ and [Fe/H] combination. 

Using the linear relation drawn through the points with [Fe/H] above --0.35, for [Fe/H] = +0.45 $\pm$0.05, the predicted $(b-y)_0$ at this H$\beta$ is 0.458 $\pm$ 0.006. The mean $b-y$ for the stars at the turnoff is 0.575 $\pm$ 0.003 (s.e.m.), leading to E$(b-y)$ = 0.117 $\pm$ 0.009 or E$(B-V)$ = 0.160 $\pm$ 0.012. With this reddening value, [Fe/H] from $hk$ would be slightly above +0.45 and the intrinsic color would be redder by less than 0.004 mag, leaving a final value of E$(b-y)$ = 0.113 and E$(B-V)$ = 0.155. Given the uncertainties in the approach due to the linear fit in Fig. 8, a higher intrinsic color and thus a slightly lower reddening estimate is plausible. However, a value as low as E$(B-V)$ = 0.09 \citep{stet} would require $(b-y)_0$ = 0.51 at [Fe/H] = +0.43, which seems outside the limits of the scatter in Fig. 8.

\section{DISCUSSION}
\subsection{The NGC 6791 $m_1$ Anomaly}

In our Phase 1 analysis of the reddening and metallicity, the metallicities derived for NGC 6791 from the $m_1$ and $hk$ indices agreed within the errors. In fact, we noted that a change of only 0.01 mag in the zero-point of the $m_1$ system would bring the two results into perfect agreement for an adopted E$(b-y)$ = 0.16. If the reddening is reduced by almost 0.05 mag, the value of $m_1$ drops by 0.015 mag and the metallicity declines by almost 0.2 dex, enhancing the discrepancy. To demonstrate the problem, in Fig. 9 we plot $m_1$ vs. $b-y$ for the same stars identified in Fig. 8, all of which have H$\beta$ between 2.596 and 2.600, inclusive. If the intrinsic color of the turnoff is  $b-y$ = 0.462, the predicted $m_1$ value is 0.280. If we take a straight average of $m_1$ for the 10 bins in Table 3, the mean $m_1$ is 0.196 $\pm$ 0.003 (s.e.m.); using a weighted average and/or excluding the brightest bin lowers the mean by less than 0.002 mag, while reducing the scatter. We will use the highest $m_1$ value to illustrate the point. With E$(b-y)$ = 0.113, E$(m_1) = -0.036$, so the reddening corrected $m_0$ is 0.232, almost 0.05 mag less than expected for the observed metallicity. This, in part, explains why the reddening estimate was so much larger than expected, $m_1$ is too small for the true metallicity and therefore, the predicted colors at the turnoff are too blue.

The low values of $m_1$ could be readily dismissed as the product of photometric errors though, as we have stated earlier, the size of the discrepancy is larger than we would concede as likely due to calibration problems. An alternative solution may exist that additionally resolves a long-standing problem with past photometric studies of NGC 6791.

One of the reasons that the argument over the true abundance of NGC 6791 has persisted for so long is that some techniques that should give reliable abundance estimates have produced [Fe/H] = 0.0 to +0.2. A perfect example is the DDO photometry  of  \citet{ja84} which, when analyzed, implied E$(B-V)$ = 0.10 and [Fe/H] = --0.08 or nearly solar. When reanalyzed with the improved calibration of \citet{tw96}, \citet{tat97} found [Fe/H] = +0.09 for E$(B-V)$ = 0.15. Clearly, the reddening issue cannot explain why the cluster appears so metal-poor compared to what now appears to be the answer, [Fe/H] above +0.4. However, the DDO metallicity indicator is, among other absorption features, dependent upon the strength of the CN band. If these elements have anomalous abundances, the DDO index for giants within a given cluster will exhibit scatter. If the [CN/Fe] ratio varies from cluster to cluster with the same [Fe/H], DDO photometry will yield different metallicities for clusters with the same [Fe/H]. This observational fact has been known and studied for decades among field halo giants and within globular clusters, where primordial and evolutionary variations in the C, N, and O abundances produce unusual scatter among the DDO indices. Spectroscopic results of the last year reveal that not only do the stars of NGC 6791 {\bf not} show alpha-enhanced elements, as originally proposed to explain the problems with the CMD's of NGC 6253 and NGC 6791, but the stars in NGC 6791 are deficient in C, with [C/Fe] between --0.2 \citep{gr06} and --0.35 \citep{or06}, and mildly deficient in O. Thus, the DDO data may well have correctly indicated all along that C within the cluster is closer to solar than Fe. Moreover, the bandpass of the DDO filter that measures the CN band is very similar to the $v$ filter in the $uvby$ system and the CN band lies within the wings of the $v$ filter.  It is therefore possible that the $m_1$ index, as observed for the turnoff stars of NGC 6791, is also correct and, as in the case of the DDO for the giants, is indicating that [C/Fe] is significantly subsolar. Note that the spectroscopic data for Ca show that the [Ca/Fe] is between +0.05 \citep{or06} and --0.03 \citep{ca06}, implying that the $hk$ index should give the correct abundance for [Fe/H].

\subsection{NGC 6253 Revisited}
As discussed in Sec. 1, the idea of resolving the discrepancy between the commonly accepted age of NGC 6791 and the derived high reddening value through the use of alpha-enhanced models originated with the comparable problem for NGC 6253. It follows that if the excessive reddening for NGC 6791 is a product of a flawed relation between [Fe/H] and $(b-y)_0$, a similar solution may apply to NGC 6253. To test this option, we made use of the $uvbyCa$H$\beta$ photometry of probable single-star members of the cluster near the turnoff, as derived in \citet{tw03}. The initial sample consisted of 99 stars bluer than $b-y$ = 0.70, including blue stragglers. Due to the younger age of NGC 6253, the turnoff region included a hydrogren-exhaustion phase hook and, below the hook, considerably more curvature at the base of the turnoff. To make the comparison with NGC 6791 more appropriate, we restricted the sample in NGC 6253 to stars along the vertical portion of the turnoff between $V$ = 14.4 and 16.6, leading to a final list of 80 stars. These 80 stars were averaged, producing mean colors of 0.564 $\pm$ 0.004, 0.193 $\pm$ 0.003, 0.404 $\pm$ 0.004, 2.624 $\pm$ 0.002, and 0.743 $\pm$ 0.006 for $b-y$, $m_1$, $c_1$, H$\beta$, and $hk$, respectively. The errors listed are the standard errors of the mean.

From our spectroscopic catalog of stars with $uvby$ and H$\beta$ photometry, we selected 42 stars in the range H$\beta$ = 2.622 to 2.626, inclusive. The plot of $b-y$ as a function of [Fe/H] for these 42 stars is illustrated in Fig. 10.  An obvious weakness of this figure in comparison with the turnoff of NGC 6791 is that only one star has [Fe/H] above +0.2 and no stars with [Fe/H] above +0.35 are included.  We have superposed the trend derived from the cooler stars discussed for NGC 6791, merely shifting the curve down in $b-y$ at a given [Fe/H] by 0.04 mag.

To estimate the intrinsic color of the turnoff, we need [Fe/H]. Following the same procedure adopted for NGC 6791, the mean photometric indices for NGC 6253 have been processed through the same H$\beta$-based metallicity calibration relations as NGC 6791 under a variety of assumptions for E$(b-y)$. As E$(b-y)$ varies from 0.15 to 0.12 to 0.09, [Fe/H] from $hk$ declines from 0.59 to 0.58 to 0.56; the comparable change in [Fe/H] based upon $m_1$ is 0.62, 0.52, and 0.42. If we assume [Fe/H] = 0.58, the linear relation on Fig. 10 predicts $(b-y)_0$ = 0.435, leading to E$(b-y)$ = 0.129, virtually identical to the value of 0.12 used to derive the metallicity. A closer look at Fig. 10 shows that if one ignores the line based upon NGC 6791, an extrapolation of the trend from the data to between [Fe/H] = 0.55 and 0.60 could produce a plausible value as high as $(b-y)$ = 0.45. Thus, our best estimate of the reddening of the cluster is E$(b-y)$ = 0.120 $\pm$ 0.015  (E$(B-V)$ = 0.16 $\pm$ 0.02) with a metallicity of [Fe/H] = +0.58 $\pm$ 0.01, on the same scale where NGC 6791 has [Fe/H] = +0.45. The biggest uncertainty in the reddening estimate remains the absence of nearby field stars with metallicities and temperatures in the same range as the turnoff of NGC 6253. 
It should be pointed out that the reddening in the direction of NGC 6253 from the maps of \citet{sc98} is E$(B-V)$ = 0.35. However, unlike NGC 6791 which lies 800 pc above the galactic plane, NGC 6253 is located between 170 and 210 pc below the plane, depending upon the adopted distance modulus. Thus, the fraction of the full reddening along the line of sight that affects the cluster should be significantly smaller than the full effect assumed for NGC 6791.

Before continuing, two points should be noted. First, the current estimate for [Fe/H] from $hk$ is only slightly lower than the original H$\beta$-based value of +0.68. The decline is a combination of three factors: the reddening value adopted here is smaller by 0.07 mag, the slope of the metallicity calibration has been slightly revised since the original discussion of NGC 6253, and no adjustment has been made to the photometry to account for potential evolutionary effects on the $hk$ and $m_1$ indices, as was done earlier; if we include the latter, the metallicity rises by 0.03 dex. Second, the metallicity from $m_1$ at the derived reddening is in excellent agreement with the $hk$ value, in contrast with NGC 6791 where the [Fe/H] is too low by a statistically significant amount. If our solution to the $m_1$ anomaly in NGC 6791 is correct and the anomaly does not exist in NGC 6253, this may be interpreted as an indication that the [CN/Fe] ratio in NGC 6253 is closer to solar.  DDO photometry of the giants and/or spectroscopic CN abundances might shed some light on this interpretation.

With a revised reddening and a more definitive estimate of the metallicity of NGC 6253, derivation of the age and distance is now possible. For isochrones, we use only the scaled-solar set from \citet{pad} with (Y,Z) = (0.39, 0.07) or [Fe/H] = +0.57, virtually identical to the derived value. The results for the data from \citet{tw03} superposed upon the isochrones for ages 2.0, 2.5  and 3.2 Gyr, adjusted for E$(B-V)$ = 0.16  and an apparent modulus of $(m-M)$ = 11.9, are illustrated in Fig. 11. The morphology of the turnoff is a good match to the models for an age of 2.5 $\pm$ 0.3 Gyr; the slope of the subgiant branch is excellent. The weakness in the fit arises in the giant branch, where the observed stars are significantly bluer than predicted by typically 0.1 to 0.15 mag. Moreover, the observed clump is slightly brighter than predicted by about 0.1 mag. The color difference, while annoying, is less of an issue than it appears given the uncertainties in the temperatures for the models of the giant branches and  the conversions from the theoretical plane to the observational, a problem we will return to in the next section. If the offsets discussed for these isochrones in the next section are appropriate, the age of the cluster should be increased to about 3 Gyr and the apparent modulus potentially reduced to 11.65. What is definitely true is that the revised reddening destroys the consistency of the cluster match to alpha-enhanced isochrones with [Fe/H] = +0.32, the optimal solution in the earlier analysis. An optimal match at the derived reddening of E$(B-V)$ = 0.16 is obtained with scaled-solar isochrones with [Fe/H] = +0.32, $(m-M)$ = 11.75, and an age of 3.5 $\pm$ 0.3 Gyr, but the improved quality of this fit is dominated by the need to match the color of the giant branch.

\subsection{NGC 6791: Age, Distance, and Galactic Context}
One rationale for detailed studies of clusters like NGC 6791 and NGC 6253 is to place them within the context of Galactic evolution. To do so requires reliable and internally consistent estimates of the reddening, metallicity, age, and distance. It has already been pointed out that recent spectroscopic data have shown that NGC 6791 is super-metal-rich, though just how rich depends on the study under discussion. Since we now have an internally consistent reddening  (E$(B-V)$ = 0.155) and metallicity ([Fe/H] = +0.45) estimate from $hk$ photometry, do the spectroscopic values agree at the appropriate level for this assumed reddening?

Comparisons for two of the three studies are straightforward. \citet{gr06} adopted E$(B-V)$ = 0.15 in defining the temperature scale to derive [Fe/H] = +0.47 $\pm$0.09 (internal and external errors included) from four red clump stars observed at S/N between 40 and 85. By pure coincidence, boosting the reddening slightly will lower the derived [Fe/H] and make the metallicity identical within the uncertainties to the $hk$ value. \citet{or06} adopted the reddening values from \citet{sc98}, which imply E$(B-V)$ = 0.154, producing [Fe/H] = +0.35 $\pm$ 0.1 (internal and external errors combined) from infrared spectra of six M giants. The importance of these results lies in what happens if we significantly lower the reddening to E$(B-V)$  = 0.09, the value adopted in the third spectroscopic study by \citet{ca06}. For \citet{gr06}, this is equivalent to lowering the temperature scale by $\sim$150 K which, using the scaling from their discussion, should $raise$ [Fe/H] by over 0.1 dex. In contrast, adopting the same lower reddening in \citet{or06} $lowers$ their final abundance by more than 0.1 dex. Finally, \citet{ca06} adopt E$(B-V)$ = 0.09 to define the temperature scale and obtain [Fe/H] = +0.39 $\pm$ 0.01 (internal errors alone) from 10 giants. Unfortunately, no information is provided regarding the impact of varying the assumed atmospheric parameters on the final abundance estimate. However, the spectral region covered by \citet{ca06} is included in the analysis of \citet{gr06} and both studies derive abundances through comparisons with synthetic lines. If we adopt the same effect found in \citet{gr06}, raising the reddening from E$(B-V)$ = 0.09 to 0.15 would decrease the derived [Fe/H] from \citet{ca06} to about +0.28. 

Given the apparent [Fe/H] of +0.45, what does an increase in E$(B-V)$ do to the derived age? The most recent discussion of the cluster age is that of \citet{ca06} where it is concluded that an adopted value of E$(B-V)$ = 0.09 $\pm$0.01 requires an age of  8 $\pm$1 Gyr when the cluster is optimally matched simultaneously to the $BV$ and $VI$ isochrones of \citet{pad} for [Fe/H] = +0.39. If one raises the metallicity to +0.45 and increases the reddening by 0.06 mag, the age must be reduced to less than 7 Gyr. Since the increased reddening has a greater impact than the small change in metallicity, it is important to understand why E$(B-V)$ = 0.09 is set as the accepted value. The fundamental claim is that this estimate has been, within the errors, the consistently derived value from studies over the last few years. In particular, it is, within the errors, the value predicted by \citet{ch99} (E$(B-V)$ = 0.10 $\pm$ 0.01 for [Fe/H] = +0.40 and an age of 8 Gyr ), by \citet{stet} (E$(B-V)$ = 0.09 for [Fe/H] = +0.40 and an age of 12 Gyr), and by \citet{ca05} (E$(B-V)$ = 0.14 $\pm$ 0.04 for [Fe/H] = +0.4 $\pm$ 0.1 and an age between 7.5 and 9 Gyr). The problem is that the analysis by \citet{ca06} reaches the conclusion that the first two studies are in error, but makes no adjustment to the fact that both almost certainly underestimate the reddening. The analysis of \citet{ch99} derives the reddening and age for various assumed values of the metallicity through comparison of the photometry of  \citet{mj94} and \citet{kr95} to an older version of the Yale isochrones \citep{gu89,gu92}. Using the comprehensive data set from \citet{stet} and the revised Yale (hereinafter referred to as the $Y^2$) isochrones \citep{yi01}, \citet{ca06} derive a best match between theory and observation for an age between 8 and 9 Gyr and E$(B-V)$ = 0.13, superseding the earlier work of \citet{ch99}. Using the $Y^2$ isochrones in the infrared \citep{yi03}, \citet{ca05} conclude that the optimum match for [Fe/H] = +0.4 requires E$(B-V)$ = 0.13. The slightly higher estimate of 0.14 is tied to the colors of the horizontal branch in clusters, not the isochrones.

For \citet{stet}, the discordance is even greater. \citet{ca06} point out that while E$(B-V)$ = 0.09 for \citet{stet}, the derived age is 12 Gyr, but state that the large age is due to an anomalously small distance modulus, implying a fundamental error in the calibration of the isochrones used to derive the distance and age. If, however, the isochrones are seriously flawed, there is no reason to believe they will generate reliable colors or consistent reddening estimates. Moreover, the age and reddening are not the result of an anomalously low distance modulus. The optimal combination of age and metallicity are selected using differentials in color and magnitude relative to the turnoff point, differentials that are not dependent upon an assumed modulus. Given a plausible range of age and metallicity, one can then use the two-color diagram of the turnoff in $(B-V)$ and $(V-I)$ to define a unique combination of reddening, metallicity, and age (see Fig. 22 of \citet{stet}.) Thus, independent of the distance, the isochrones predict that the turnoff colors require [Fe/H] = +0.4, E$(B-V)$ = 0.09 and an age of 12 Gyr.

The fundamental point is that, even if [Fe/H] is well-determined, simultaneous solutions for the reddening and age through isochrone matches with multicolor CMD's ($B-V$, $V-I$, $B-I$, $J-K$) are only as reliable as the isochrones in the various planes. \citet{ca05} and \citet{ca06} produce a reddening of 0.13 or higher when using the $Y^2$ isochrones, but \citet{ca06} derives a reddening of 0.09 using the \citet{pad} models; all of these are incompatible with the age and reddening combinations defined by \citet{stet}. Note: the same issue holds for comparing the colors of stars in NGC 6791 with standard colors for nearby stars and clusters - the accuracy is only as good as our assumption that some of the nearby stars are analogs to those in NGC 6791.  It is crucial to recognize the fact that the only reddening estimate that does not depend upon the intrinsic properties of the stars in the cluster is that of \citet{sc98}, which leads to E$(B-V)$ = 0.154.

If we use only the $B-V$ color of the turnoff, adopt E$(B-V)$ = 0.15 and [Fe/H] = +0.45, what age and distance modulus are generated by the currently available isochrones? Rather than adding another layer of isochrone comparisons to the already extensive literature on this cluster, we will refer to the most recent analyses and differentially adjust the results for the required changes in reddening and metallicity. The first isochrone adjustment is for the $Y^2$ match as delineated in Fig. 7 of \citet{ca06}. With $E(B-V)$ = 0.13 and [Fe/H] = 0.39, the optimal match produces an age of 8.5 $\pm$ 0.5 Gyr and an apparent modulus of 13.35, though this should be regarded as a lower limit to the modulus given the increasing discrepancy between the isochrones and observations at fainter apparent magnitude. An increase of only 0.02 in reddening will lower the age at this [Fe/H] by approximately 0.9 Gyr to 7.6 $\pm$ 0.5 Gyr. The apparent modulus will increase by about 0.12 mag.  If we make the additional demand that the isochrone metallicity be increased from +0.39 to +0.45, the age is reduced to approximately 6.9 Gyr and the apparent modulus is raised to $(m-M)$ = 13.55 (rounded to the nearest 0.05 mag). 

For the isochrones discussed by \citet{stet} as quantified in their Fig. 22, with [Fe/H] = +0.40 and E$(B-V)$ = 0.15, the cluster turnoff in $B-V$ implies an age of 8.2 Gyr. Their apparent modulus of 13.1 is boosted to 13.45, effectively identical to that of the $Y^2$ comparison. Finally, if we additionally boost [Fe/H] to +0.45, the age is reduced to 7.6 Gyr and the apparent modulus is increased to 13.50.  

The last set of isochrones to be probed is that of \citet{pad} as presented by \citet{ca06}. As shown in Fig. 5 in \citet{ca06}, with E$(B-V)$ = 0.09 and [Fe/H] = +0.39, the optimal fit to the isochrones implies an age of 9.5 $\pm$ 0.5 Gyr and an apparent modulus of 13.35, though again this may be an overestimate depending on how much weight one gives to the fainter main sequence. If we raise E$(B-V)$ to 0.15, the age reduces to 6.6 Gyr and the apparent modulus increases to 13.70. If we make the additional boost to [Fe/H] = +0.45, the age lowers to less than 6 Gyr and $(m-M)$ becomes 13.75. 

What this implies is that the current uncertainties in the age and distance to NGC 6791 are dominated not by the reddening and metallicity estimates, but by the inconsistency among the isochrones. Depending on the set which one chooses, for a uniformly adopted [Fe/H] and E$(B-V)$, the age of the cluster can vary by over 1.6 Gyr and the distance by almost 0.3 mag. The fundamental issues remain the inherent differences in the model construction and  the manner in which the theoretically derived parameters of temperature and luminosity are transferred to the observational plane. While a comprehensive discussion of these complicated topics is beyond the scope of this investigation, the relevance of the latter issue can best be illustrated with the \citet{pad} isochrones that have formed the basis of our age analyses throughout this series of cluster papers. As detailed in \citet{tw99}, one can fix an absolute age, distance, and color scale for only one set of isochrones, those of solar composition, by demanding that a solar-mass star have the color and absolute magnitude of the Sun at the adopted age of the Sun. When applied to the \citet{pad} isochrones, this required an adjustment to the isochrone $M_V$ and $B-V$ of $-0.04$ mag and $-0.032$ mag, respectively, i.e., the isochrones are too red. Lowering the adopted solar color below $B-V$ = 0.65 requires a correspondingly larger correction. When tested at [Fe/H] near $-0.4$, the unevolved main sequence superposes upon the observational data with no adjustment. Note that while this leads to correct distance estimates, it does not ensure a correct age scale since there may be compensating changes in both $M_V$ and $B-V$ that lead to superposition of the main sequences but assign the wrong color and $M_V$ to stars of a given mass.

What offsets, if any, are required for the \citet{pad} isochrones at high metallicity? As a simple test, we have identified 15 unevolved main sequence stars with [Fe/H] between +0.25 and +0.35 as measured by VF. Using $M_V$ based upon Hipparcos parallaxes and $B-V$ colors as listed on SIMBAD, comparison with the \citet{pad} isochrones for [Fe/H] = +0.32 implies a significant offset between theory and observation. Depending on how one does the comparison, the isochrones are typically too bright by 0.25 mag or too red by 0.04 mag in $B-V$. In reality, there is no way at present to tell how these offsets are distributed between color and magnitude or determine if they are applicable at even higher metallicity. However, if we adopt the shift completely within color and apply it to the \citet{pad} isochrones within \citet{ca06}, the derived parameters for the cluster would be E$(B-V)$ = 0.13 at [Fe/H] = +0.39 and $(m-M)$ = 13.35. Making the final adjustment to E$(B-V)$ = 0.15 and [Fe/H] = +0.45 gives an age of  7.8 Gyr for $(m-M)$ = 13.55. Taking into account the variation among the isochrones and the modest uncertainty in the reddening, our best estimates of the cluster parameters to date are $(m-M)$ = 13.6 $\pm$ 0.1 and an age of 7 $\pm$ 1 Gyr.

To close this discussion of NGC 6791, we raise the issue of where the cluster fits within the context of Galactic evolution. The short answer is that it doesn't, at least in the sense of forming within the disk anywhere near the solar circle. Though there was some hope when metallicity estimates hovered between [Fe/H] = 0.0 and +0.2 that it might represent the high metallicity tail of the old disk distribution, at [Fe/H] = +0.45 with an age greater than NGC 188, this no longer seems plausible. The long-standing alternative tied to the eccentric orbit of the cluster is that it originated near the Galactic center where the metallicity rose higher more quickly and is simply passing through the region of the solar circle. A third possibility that has gained credibility as evidence for mergers and stellar streams grows is that NGC 6791 formed as part of an external system that has since merged with the disk and therefore tells us nothing about chemical evolution of the disk. For a recent discussion of these options, see \citet{ca06}.

Note, however, that the relative merits of the three options above are strongly dependent upon the care with which the data are treated, specifically the cluster data, as exemplified by the analysis of \citet{ca06}. How anomalous is NGC 6791? Based upon the data available to date for NGC 6253, including the reanalysis above, it is apparent there are at least two older open clusters with metallicities twice solar or higher. The fact that NGC 6253 was not included in the moderate-resolution spectroscopic survey by \citet{fj02} doesn't allow one to exclude it from a discussion of the metal-rich end of the cluster distribution.  Second, defining the extent of NGC 6791's anomaly requires comparison to a sample on the same metallicity scale. The [Fe/H]  of +0.19 for NGC 6791 in the original survey by \citet{fj} and the value of +0.11 from \citet{fj02} are often used to downplay the reliability of the technique, especially at the metal-rich end, without recognizing that the abundance scale of the final sample does not have the same zero-point as most high-dispersion spectroscopic surveys. As pointed out in past papers in this series, even a quick glance at the final data of \citet{fj02} indicates that a problem exists with at least the zero-point of the scale. Of 39 clusters, only NGC 6791 at [Fe/H] = +0.11 has a metallicity of solar or above; 37 clusters have [Fe/H] = --0.10 or lower. \citet{tat97} showed that the 62 clusters interior to $R_{GC}$ = 10 kpc on a common metallicity scale tied to the field giants of the solar neighborhood have a mean metallicity of solar.

The overall calibration of the cluster data by \citet{fj02} made use of a mixture of field giants and clusters. The metallicity scale of the field stars was based upon abundances defined using a mixture of high dispersion spectroscopy and DDO photometry. For the four clusters, the abundances were fixed at [Fe/H] = --0.1, --0.24, --0.42, and --0.79 for M67, NGC 7789, NGC 2420 and M71, respectively. The final calibration produces abundances indicating that the clusters and field stars are on the same system; the final open cluster values are --0.15, --0.24, and --0.38 for M67, NGC 7789, and NGC 2420, respectively. Recent high dispersion spectroscopic work places the abundances of M67 and NGC 7789 closer to 0.00 and --0.05 , while DDO and the $hk$ photometry predict [Fe/H] = --0.32 for NGC 2420 (see the discussions in \citet{ta05, ra06, at06}). This indicates that the \citet{fj02} scale is systematically off by 0.15 to 0.20 dex near solar metallicity. Whether the size of the correction is greater at higher [Fe/H] remains unknown, but application of this offset to NGC 6791 leads to [Fe/H]  between +0.26 and +0.31, at minimum. Thus, on a uniform scale, NGC 6791 lies three to four sigma (0.1 dex) above the mean for the open cluster distribution in the solar circle.

Finally, to repeat a regular theme of this series, for clusters interior to $R_{GC}$ = 10 kpc on a scale where the sun is at 8.5 kpc, there is no evidence for a significant metallicity gradient in the disk. The regular claim for a slope of --0.05 to --0.07 dex/kpc only occurs if one includes the cluster sample beyond $R_{GC}$ = 10 kpc. Therefore, extrapolating such a gradient from [Fe/H] = 0 at the sun to the Galactic center to predict a typical [Fe/H] near +0.4 to +0.6 in that distant region is not only incorrect, it conflicts with the direct observation of the Galactic center where the mean metallicity has been found to be comparable to that of the solar neighborhood \citep{fu06}.

\section{SUMMARY AND CONCLUSIONS}
$vby$H$\beta$ CCD photometry of the metal-rich open cluster, NGC 6791, has been presented and analyzed. If taken at face value and analyzed in the same fashion as stars and clusters of solar metallicity or less, the cluster attains an [Fe/H] between +0.37 and +0.50 from both the $m_1$ and $hk$ indices. The coupled result is a reddening value of E$(B-V)$ = 0.22 which, when applied to the CMD, requires a cluster age of less than 4 Gyr in comparison with traditional scaled-solar models for what is, at best, a flawed morphological fit between observation and theory. One can obtain a high quality fit for the $V$, $B-V$ isochrones at an acceptable age if alpha-enhanced models are adopted, but the high reddening value is incompatible with the $V$, $V-I$ data. Moreover, recent spectroscopic results indicate that NGC 6791 does, in fact, have scaled-solar abundances, with the exception of C, which is deficient relative to Fe by about a factor of two. The solution to this conundrum is of relevance to not only NGC 6791, but also to NGC 6253, the only open cluster with a metallicity comparable to NGC 6791 which exhibits a virtually identical reddening/CMD morphology problem.

The apparent solution appears to be that the traditional $(b-y)_0$-temperature-metallicity relations underestimate the $(b-y)_0$ color for stars at the very metal-rich end of the scale, leading to an overestimate for the reddening. The relations have been recalibrated using stars in the solar neighborhood with the same H$\beta$ indices as the turnoffs of NGC 6791 and NGC 6253 over a range in [Fe/H], implying E$(B-V)$ = 0.155 and E$(B-V)$ = 0.16 for the two clusters, respectively, and, from the $hk$ index, [Fe/H] = +0.45 and +0.58. We emphasize that the [Fe/H] values are extremely well determined due to the exceptionally low sensitivity to changes in E$(B-V)$. The combined uncertainty in [Fe/H] from the photometric zero-points and the reddening is below $\pm$0.03 dex.

In contrast, the reddening estimates are tied to an assumed linear relation between $(b-y)_0$ and [Fe/H] at a given H$\beta$. For the turnoff temperature of NGC 6791, there are stars in the solar neighborhood with [Fe/H] comparable to the cluster. At the turnoff temperature of NGC 6253, one is forced to extrapolate from a sample that reaches a metallicity only half that of the cluster. Clearly the uncertainty in the NGC 6253 reddening is larger than for NGC 6791 for which the combined internal and external errors in E$(B-V)$ lead to an uncertainty of $\pm$0.016. 

An intriguing result of the revised reddening estimates is that while the metallicity estimates from $m_1$ and $hk$ are the same within the uncertainties for NGC 6253, the lower reddening value generates [Fe/H] from $m_1$ which is a factor of two lower than the $hk$-based result for NGC 6791.  This discrepancy may be real. The $m_1$ index includes the CN-band that is the dominant metallicity indicator among giants in the DDO system. Since it now appears that NGC 6791 is C-deficient, the possibility exists that the long-standing low abundance estimate for NGC 6791 from DDO photometry may have been correct, rather than a failure of the system to work at high metallicity, and may produce $m_1$ indices that are too small for the cluster. If this is correct, the photometry for NGC 6253 is valid, and the agreement between $hk$ and $m_1$ is not due to the higher temperature of the turnoff in NGC 6253, it may imply that [Ca/C] $\sim$ 0.0 for the cluster. Whether this also means that [Ca/Fe] = 0.0 remains to be seen. 

\acknowledgements
The data used in this project would not have been accessible without the help of Con Deliyannis and the excellent support provided by the WIYN 0.9m staff at KPNO. We're very grateful to Kyle Cudworth for maintaining our ongoing conversation about NGC 6791 and for supplying us with updated proper motion memberships. 
We appreciate the thoughtful comments of the referee generated by a careful reading of the paper.
Extensive use was made of the SIMBAD database, operating at CDS, Strasbourg, France and the WEBDA database maintained at the University of Vienna, Austria (http://www.univie.ac.at/webda). 
The cluster project has been helped by support supplied through the General Research Fund of the University of Kansas and from the Department of Physics and Astronomy.

\clearpage
\figcaption[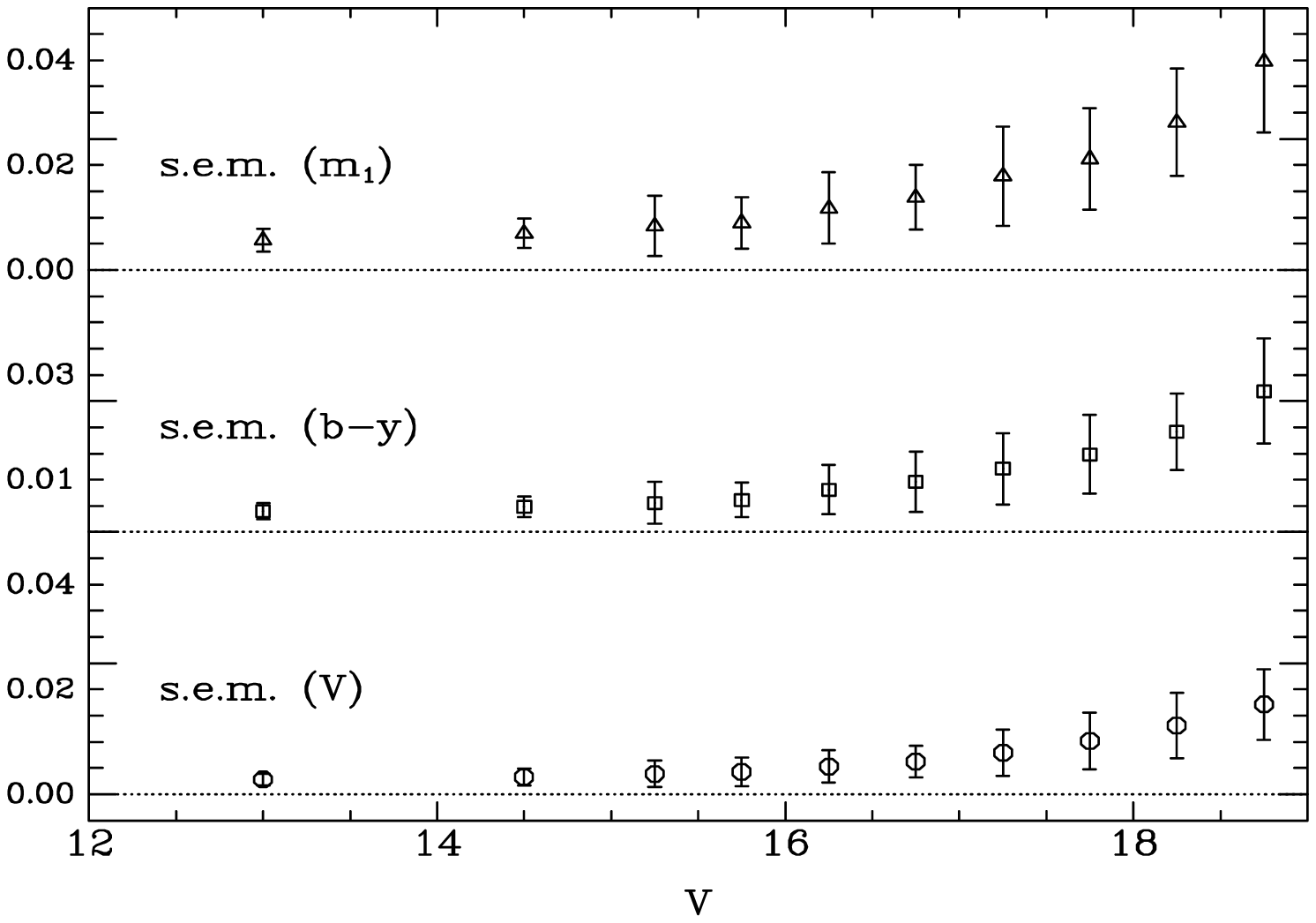]{Standard errors of the mean (sem) for the $V$, $b-y$, and $m_1$ indices as a function of $V$. Open symbols are the average sem while the error bars denote the one sigma dispersion. \label{f1}} 

\figcaption[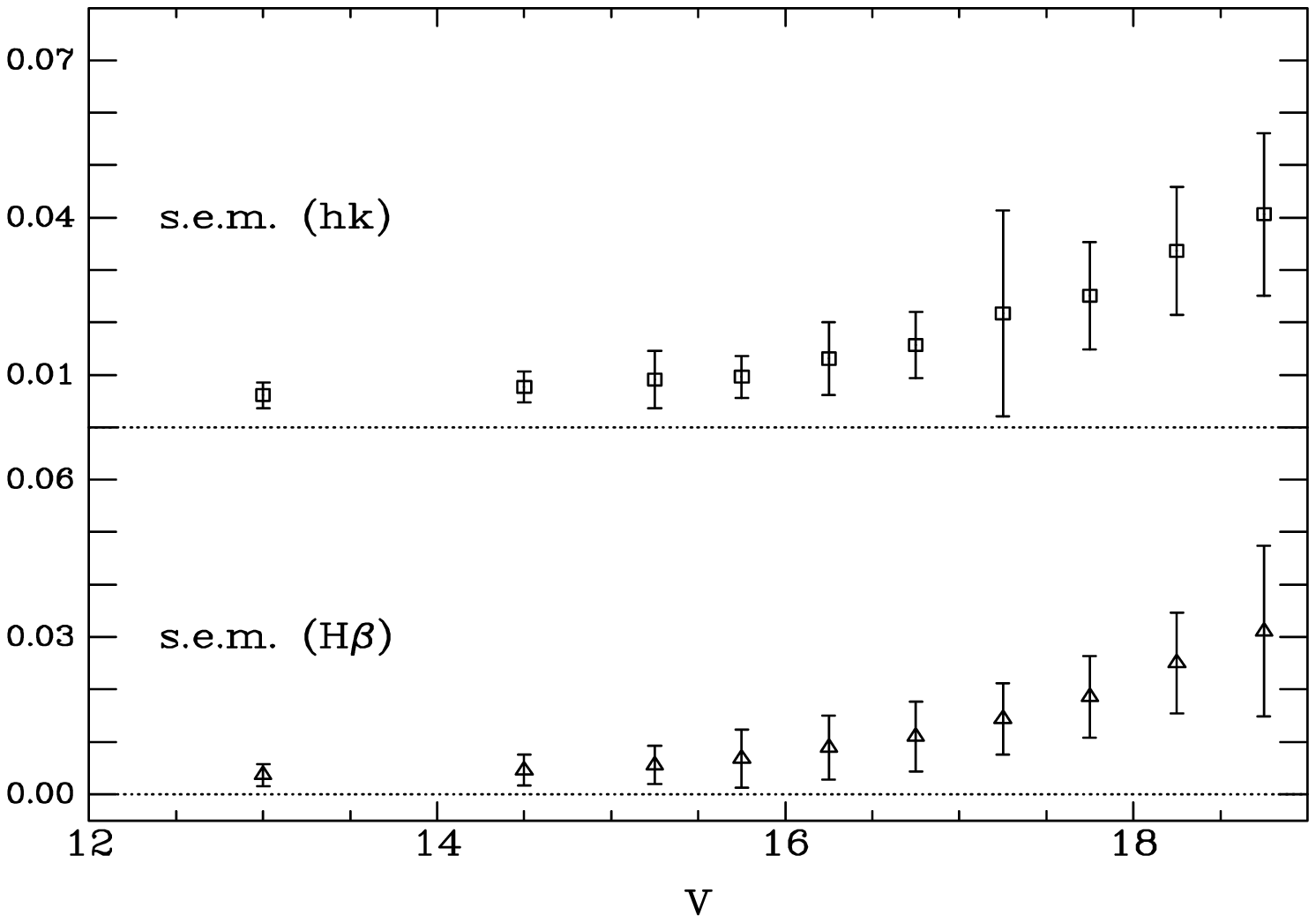]{Same as Fig. 1 for $hk$, and H$\beta$. \label{f2}}

\figcaption[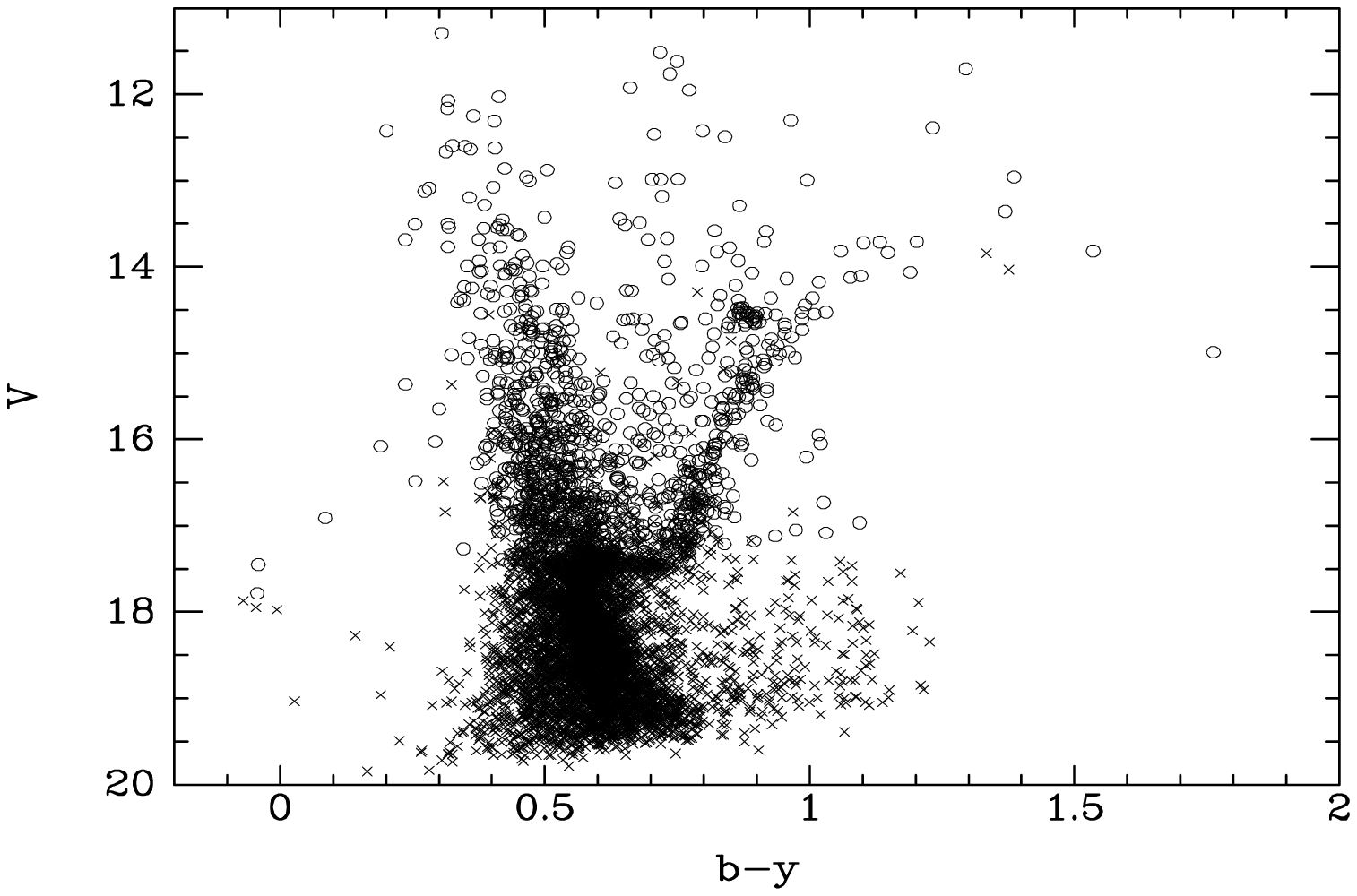]{Color-magnitude diagram for stars with at least 2 observations each in $b$ and $y$. Crosses are stars with internal errors in $b-y$ greater than 0.010 mag.\label{f3}}

\figcaption[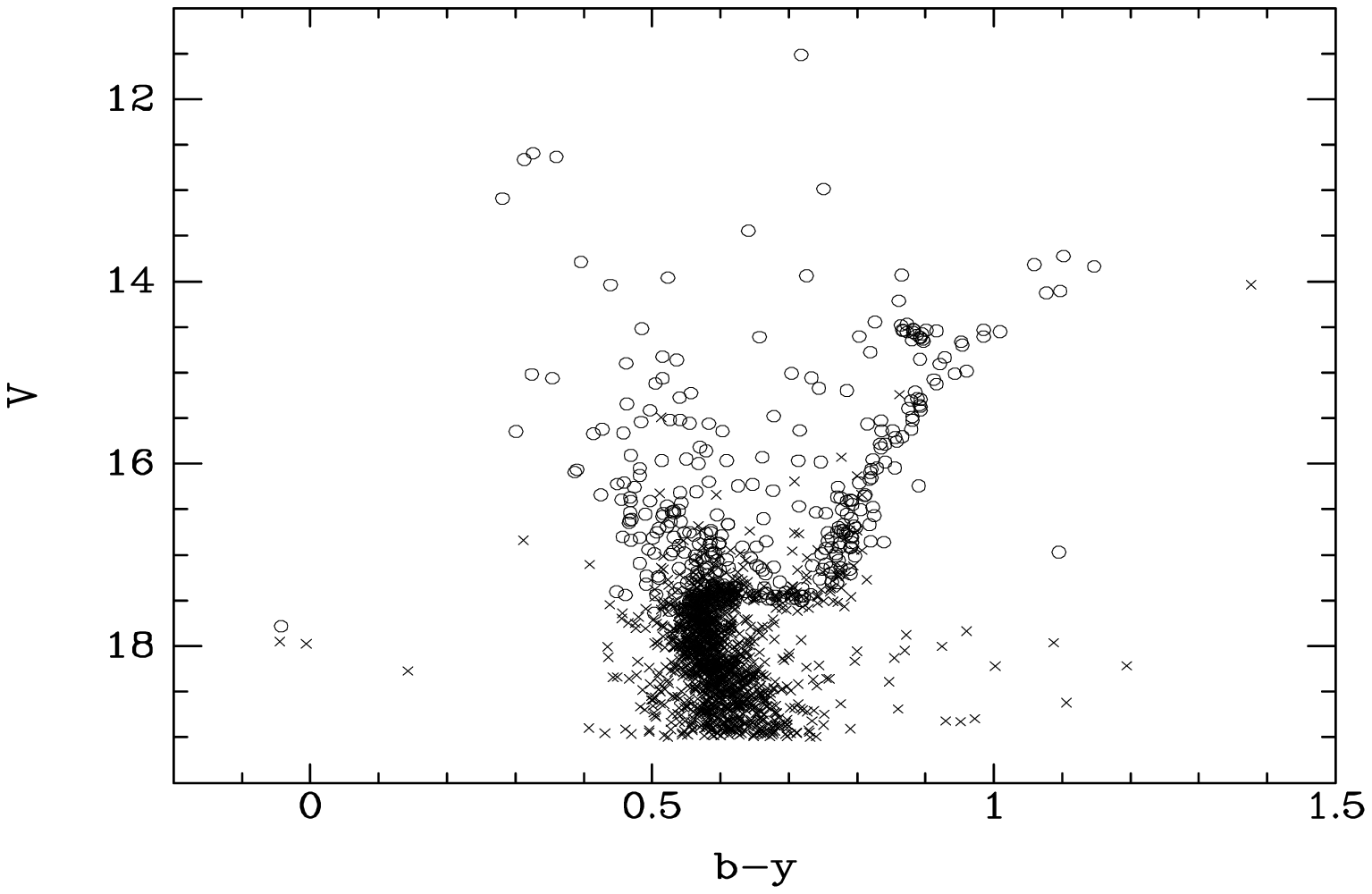]{Same as Fig. 3 for stars within 230\arcsec\ of the cluster center. \label{f4}}

\figcaption[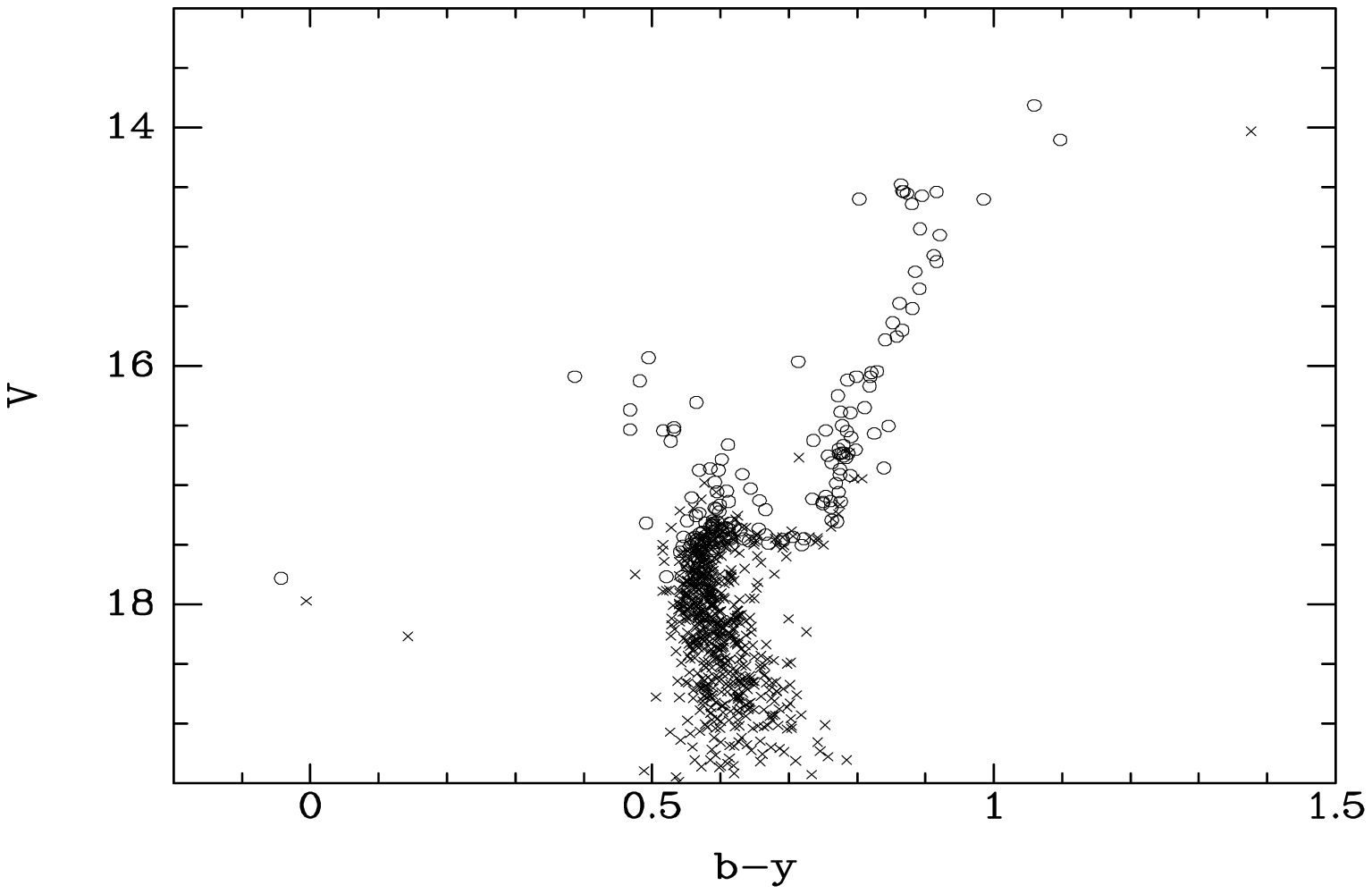]{CMD for all stars in the CCD field with membership probability of 90\% or higher. Symbols have the same meaning as in Fig. 3. \label{f.5}}

\figcaption[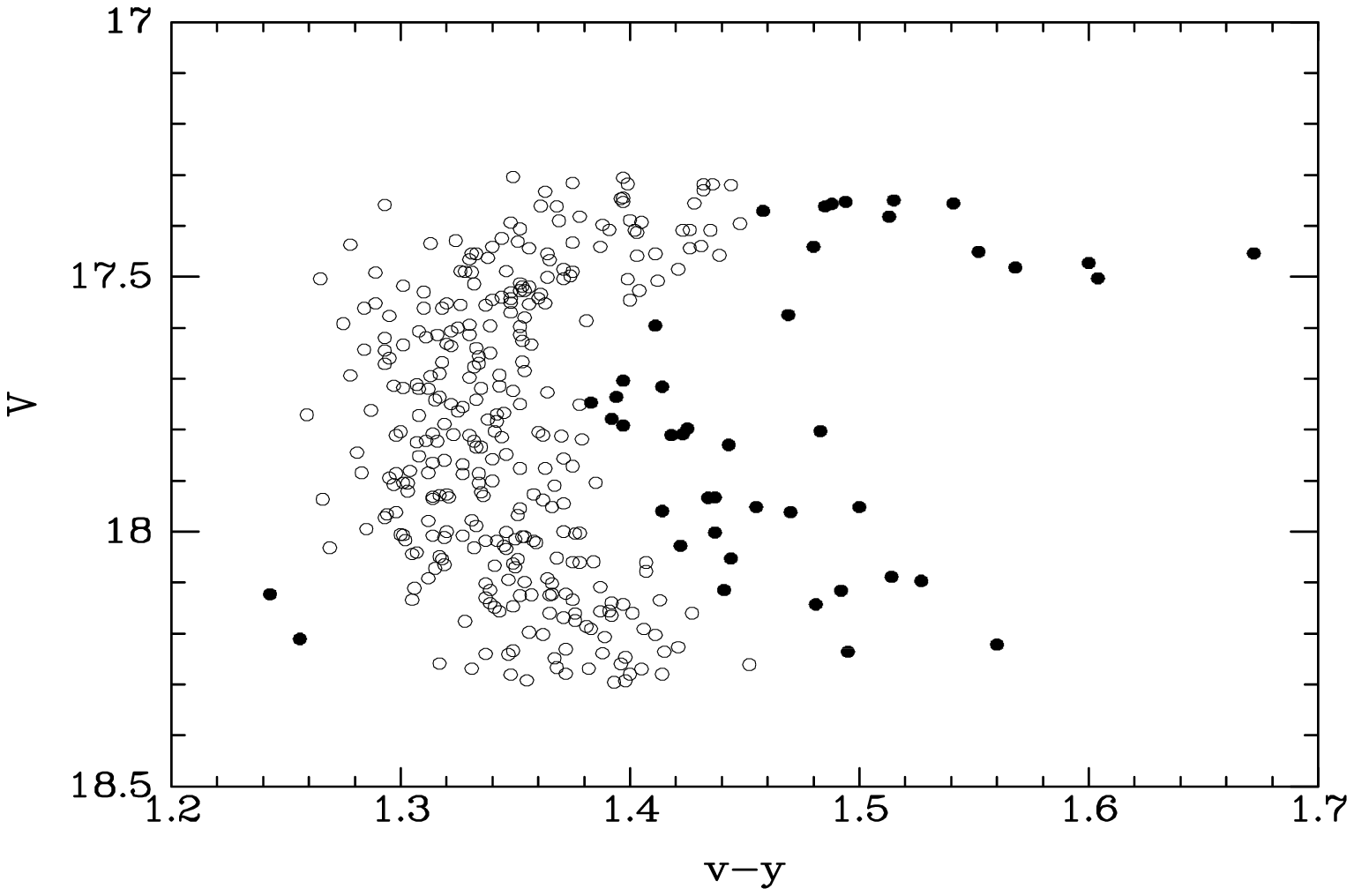]{$V,v-y$ CMD for cluster core stars at the turnoff. Filled circles are stars tagged as potential binaries, nonmembers, or photometric anomalies. \label{f6}}

\figcaption[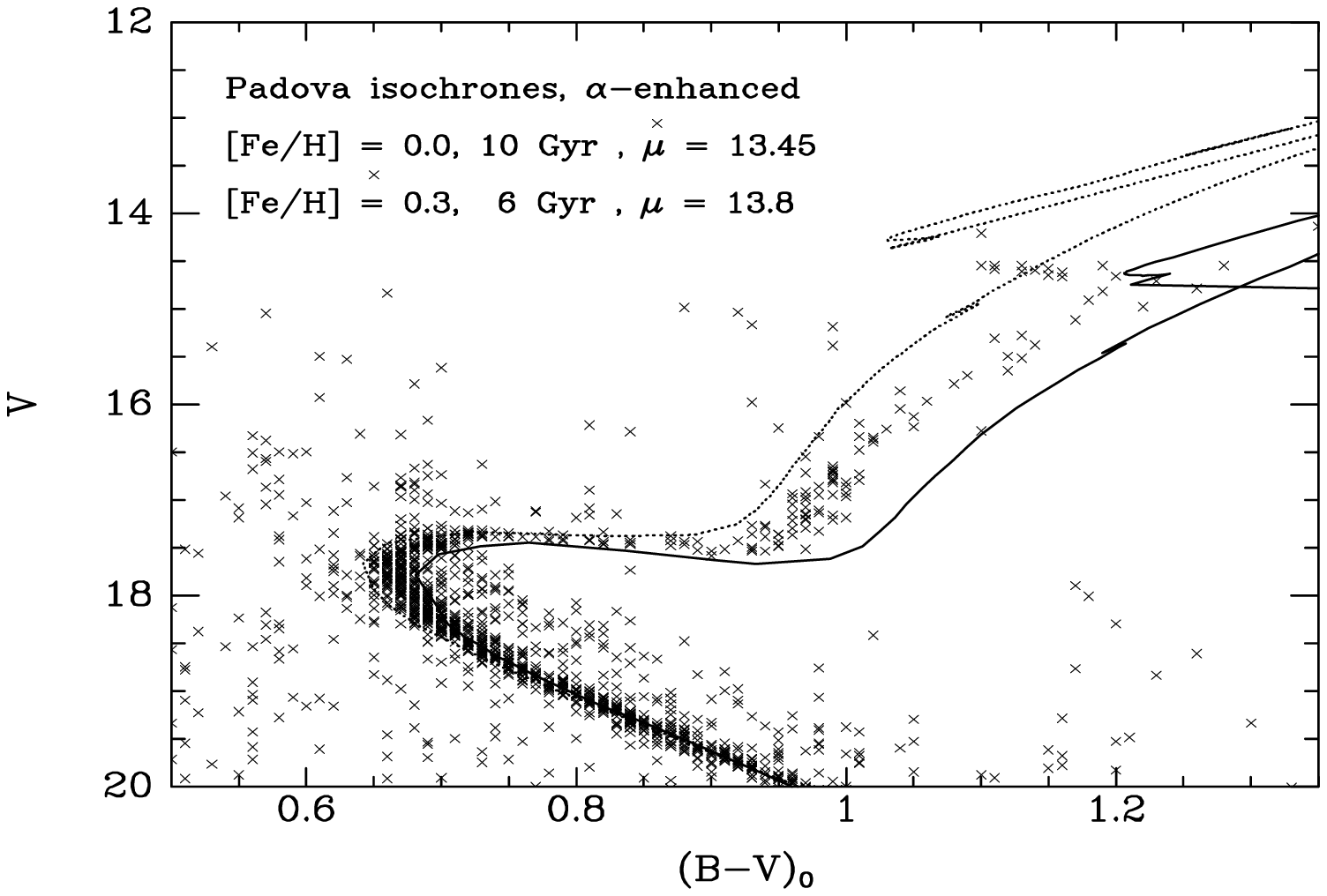]{CMD for NGC 6791 from \citet{stet} superposed upon the alpha-enhanced isochrones of \citet{pad} for [Fe/H] = 0.0 (dashed line) and +0.3. The $B-V$ colors have been corrected for E$(B-V)$ = 0.22 and the apparent moduli set to match the individual isochrones. \label{f7}}

\figcaption[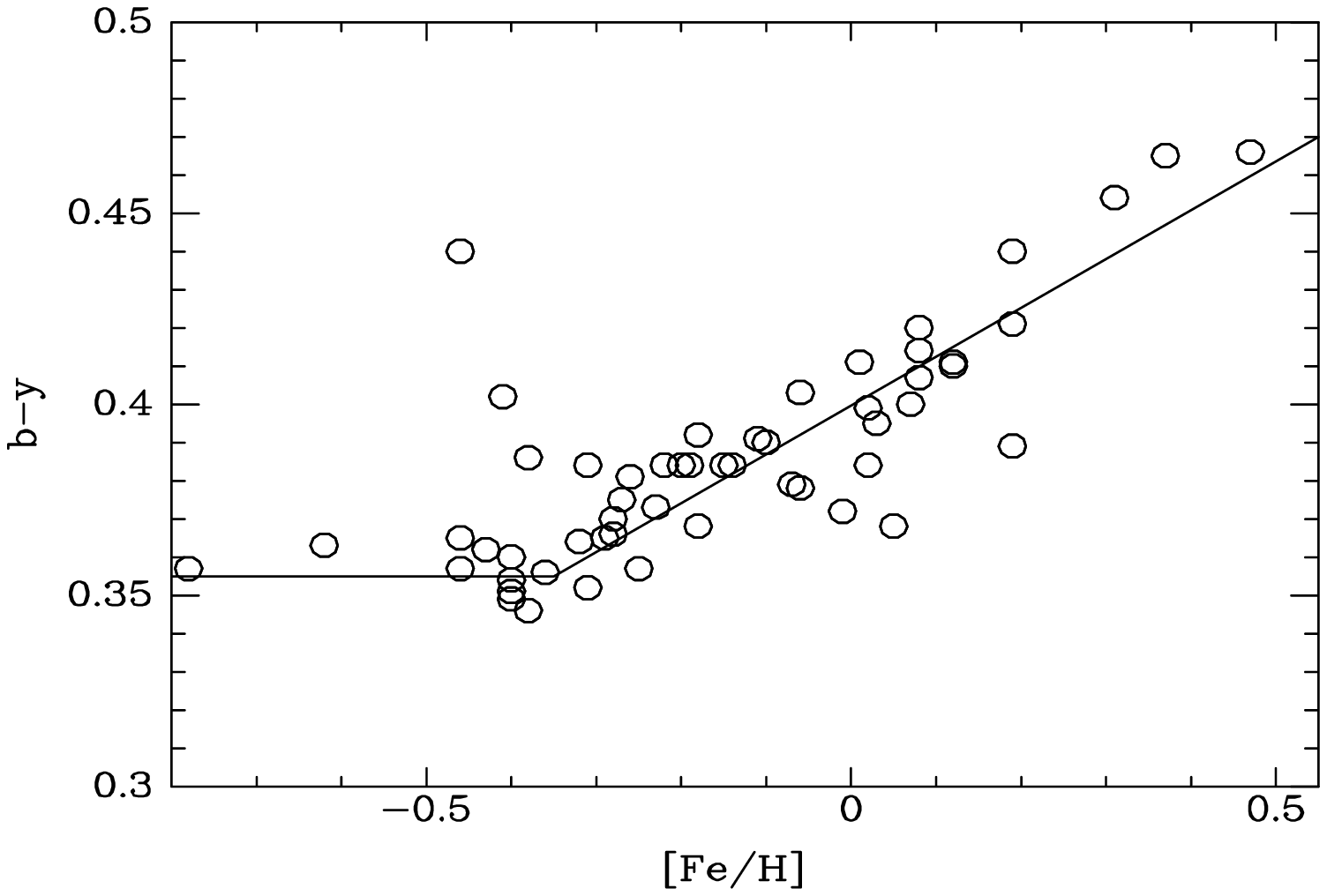]{The color dependence of $b-y$ on [Fe/H] for field stars with H$\beta$ between 2.596 and 2.600. Solid line is the derived mean relation through the points. \label{f8}}

\figcaption[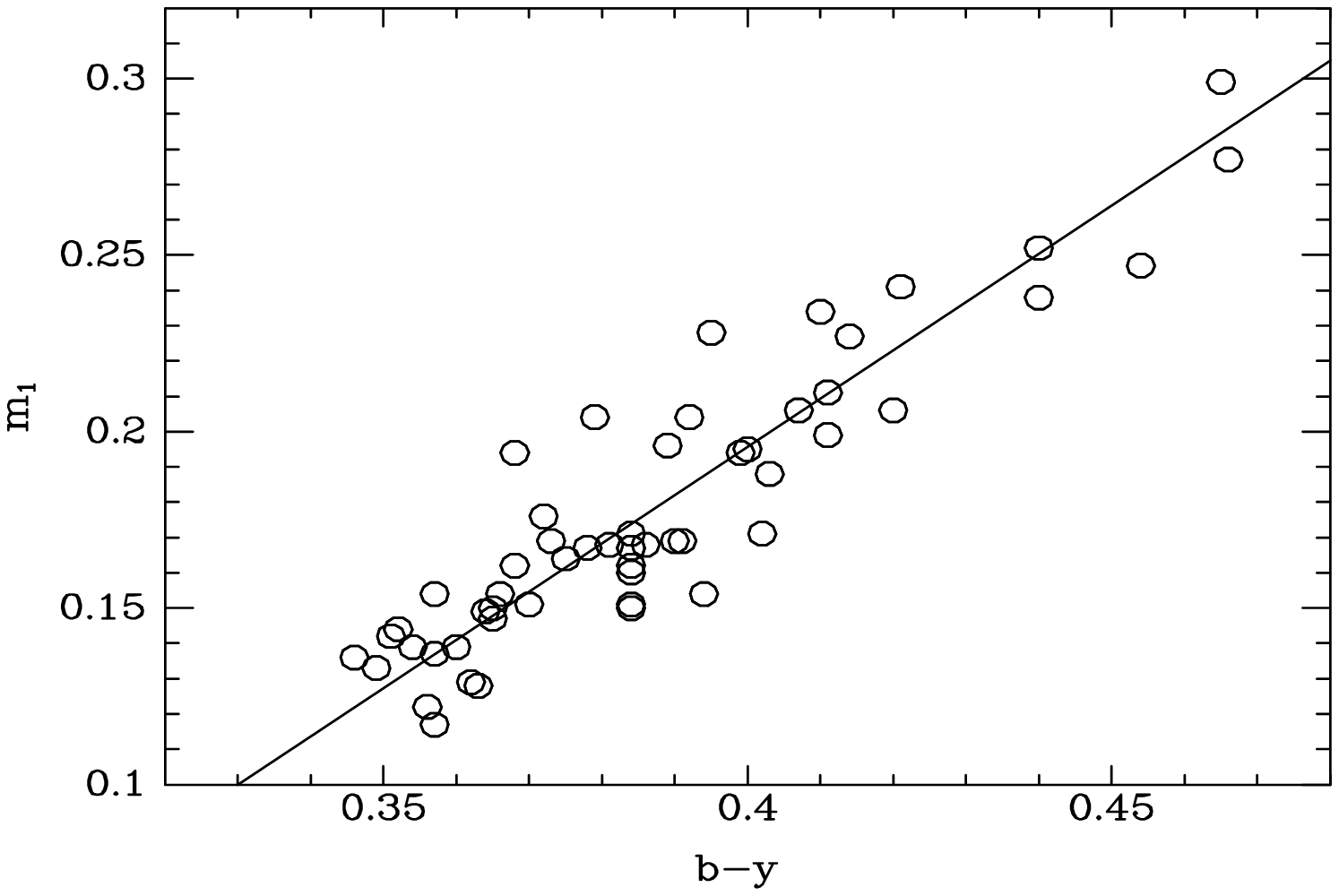]{The color dependence of $b-y$ on $m_1$ for the stars of Fig. 8. \label{f9}}

\figcaption[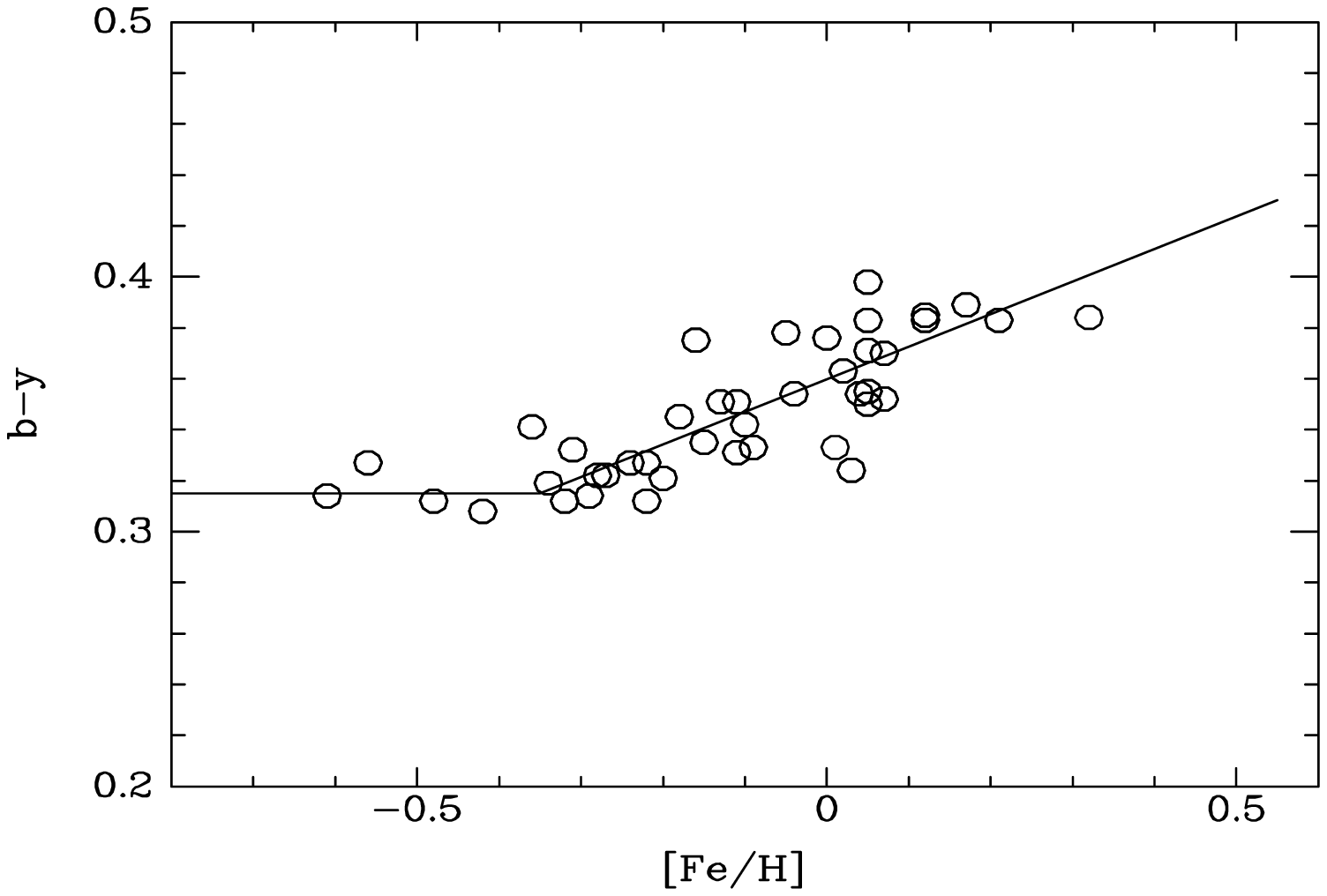]{The color dependence of $b-y$ on [Fe/H] for field stars with H$\beta$ between 2.622 and 2.626. Solid line is the derived mean relation of Fig. 8, shifted downward by 0.04 mag. \label{f10}}

\figcaption[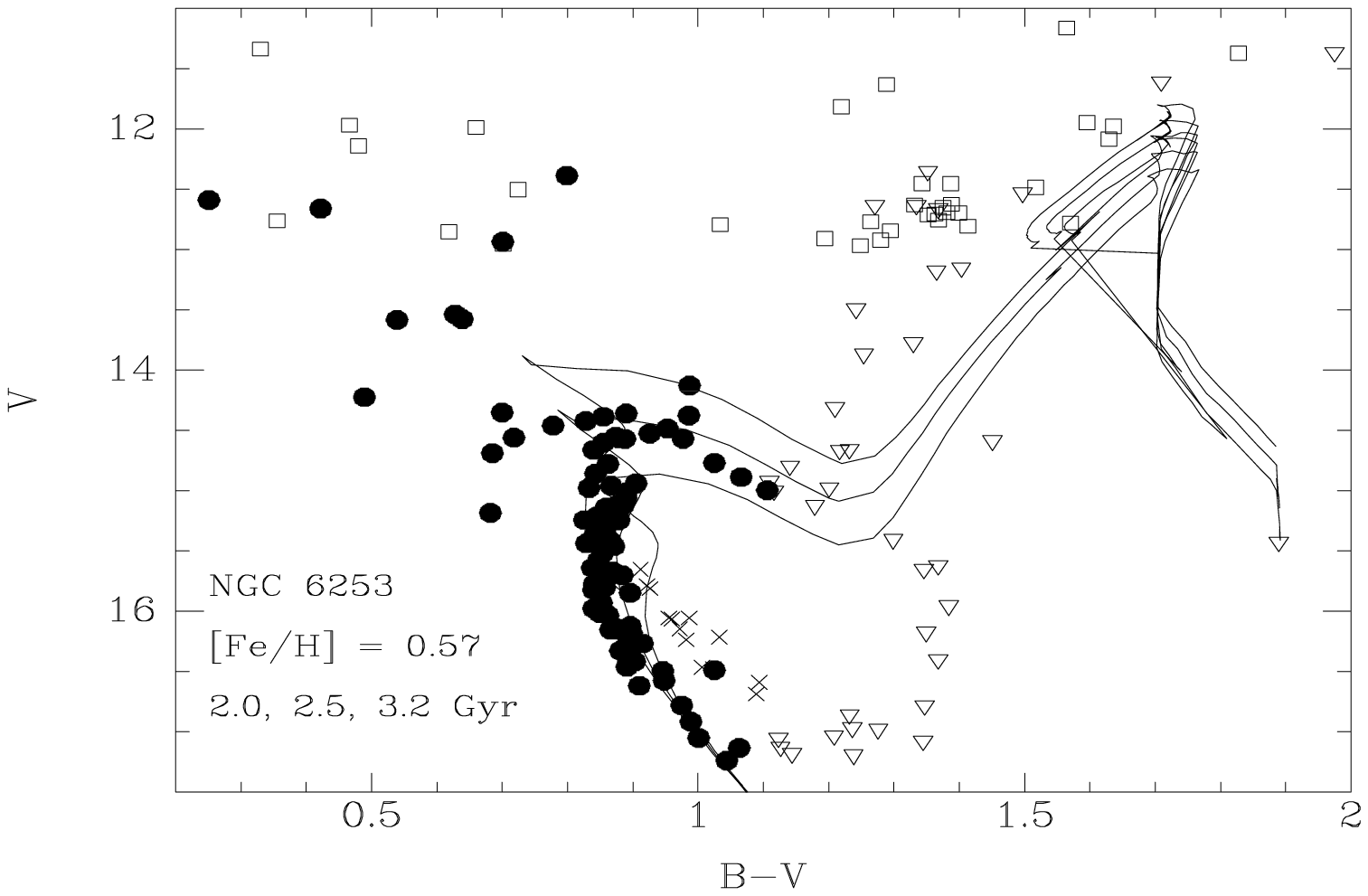]{The CMD of NGC 6253 as derived by \citet{tw03}, superposed upon the [Fe/H] = +0.57, scaled-solar isochrones of \citet{pad}. Filled circles are probable core members, crosses are probable binaries, triangles are the remaining stars within the core, and squares are all stars brighter than $V$ = 13.0. \label{f11}}

\newpage
\plotone{f1.eps}
\newpage
\plotone{f2.eps}
\newpage
\plotone{f3.eps}
\newpage
\plotone{f4.eps}
\newpage
\plotone{f5.eps}
\newpage
\plotone{f6.eps}
\newpage
\plotone{f7.eps}
\newpage
\plotone{f8.eps}
\newpage
\plotone{f9.eps}
\newpage
\plotone{f10.eps}
\newpage
\plotone{f11.eps}

\newpage
\begin{deluxetable}{lccccccccc}
\tablecolumns{10}
\tabletypesize\small
\tablewidth{0pc}
\tablecaption{Characterization of Calibration Equations}
\tablehead{
\colhead{}    &  \multicolumn{1}{c}{$V$} & 
\multicolumn{2}{c}{$b-y$} & \colhead{} & \multicolumn{3}{c}{$m_1$} &
\multicolumn{1}{c}{$hk$} & \multicolumn{1}{c}{H$\beta$} \\
\cline{3-4} \cline{6-8} \\
\colhead{} & \colhead{$All$} & \colhead{$bd/rg$}   & \colhead{$rd$} & \colhead{} &     \colhead{$bd$} &
\colhead{$rd$}    & \colhead{$rg$}   & \colhead{$All$}  & \colhead{$All$}}
\startdata
Number of photometric nights used& 5 & 5 & 2 & & 3 & 3 & 1 & 3 & 3 \\
Calibration equation slope $\alpha$ & 1.000 & 1.134 & 0.909  & & 0.828   & 1.135 & 0.945 & 1.000 & 1.140 \\
Color term $\gamma$ & 0.03\phn & \nodata  &  \nodata &  & -0.27\phn & -0.15\phn & -0.15\phn & \nodata  & \nodata \\
Typical standard deviation for & 0.025 & 0.015 & 0.008 & & 0.005 & 0.050 & 0.026 & 0.022 & 0.011 \\
\phm{word}calibration equation for one night & & & & & & & & & \\
Typical contribution to zeropoint ($\beta$)  & 0.002 & 0.001 & 0.001 & & 0.002  & 0.002 & 0.002 & 0.002 & 0.002 \\
\phm{word}s.e.m. from aperture correction & & & & & & & & &  \\
Combined s.e.m. for final & 0.003 & 0.007 & 0.013 & & 0.012 & 0.018 & 0.010 & 0.004 & 0.003 \\
\phm{word}calibration equation & & & & & & & & &  \\
\enddata
\tablecomments{Calibration equations for index $x$ are of the form $x_{std} = \alpha\  x_{instr} + \gamma (b-y)_{instr} + \beta$. 
Classes of stars include warm dwarfs {\it bd}, cool dwarfs {\it rd}, and cool giants {\it rg}.}
\end{deluxetable}

\newpage
\begin{deluxetable}{rrrrrrrrrrrrrc}
\rotate
\tabletypesize\small
\tablenum{2}
\tablecolumns{14}
\tablewidth{0pc}
\tablecaption{Extended Str\"omgren Photometry in NGC 6791}
\tablehead{
\colhead{ID}     & 
\colhead{X}     & 
\colhead{Y}     & 
\colhead{$V$}     & 
\colhead{$b-y$}     & 
\colhead{$m_1$}     & 
\colhead{H$\beta$}     & 
\colhead{$hk$}     & 
\colhead{$\sigma_V$}     & 
\colhead{$\sigma_{by}$}     & 
\colhead{$\sigma_{m1}$}     & 
\colhead{$\sigma_{\beta}$}     & 
\colhead{$\sigma_{hk}$}     & 
\colhead{$N(ybvuwnCa)$}    }
\startdata
     1 & 627.69 & 447.27 &17.871&  0.483 & 0.087 & 2.630 & 0.539 &0.008&0.015&0.022&0.017&0.024 &15,12,13,12, 9,11 \\
     2 & 627.91 & 290.97 &17.547&  0.545 & 0.198 & 2.585 & 0.792 &0.008&0.013&0.019&0.015&0.021 &15,14,13,13, 8,10 \\
     5 & 622.35 & 523.59 &18.419&  0.620 & 0.219 & 2.678 & 0.857 &0.012&0.017&0.029&0.043&0.061 & 8,10, 8, 8, 4, 2 \\
     6 & 622.18 & 419.26 &18.120&  0.715 & 0.117 & 2.651 & 0.797 &0.008&0.018&0.026&0.026&0.039 &12,10, 9, 9, 7, 2 \\
     7 & 620.82 & 592.63 &19.272&  0.697 & 0.277 &\nodata&\nodata&0.019&0.050&0.075&\nodata&\nodata& 6, 2, 1, 6, 0, 0 \\
     8 & 620.08 & 546.70 &17.444&  0.603 & 0.220 & 2.615 & 0.841 &0.005&0.008&0.013&0.018&0.015 &15,11,13,13,11, 8 \\
    11 & 615.09 & 540.33 &17.670&  0.562 & 0.230 & 2.662 & 0.856 &0.009&0.013&0.018&0.014&0.026 &13,13,13,13, 7, 8 \\
    12 & 614.91 & 440.70 &17.373&  0.452 & 0.074 & 2.598 & 0.432 &0.006&0.009&0.013&0.014&0.014 &13,14,14,13,11,13 \\
    14 & 613.14 & 434.47 &17.966&  0.955 & 0.261 & 2.659 &\nodata&0.015&0.025&0.040&0.024&\nodata& 6, 6, 5, 2, 6, 0 \\
    15 & 611.83 & 417.92 &18.080&  0.486 & 0.075 & 2.719 & 0.609 &0.009&0.013&0.018&0.014&0.021 &12,11,11,12, 6, 9 \\
       &        &        &          &    &       &       &       &     &      &    &      &      & \\
    16 & 611.57 & 239.89 &18.558&  0.420 & 0.159 & 2.598 & 0.612& 0.012&0.019&0.027&0.021&0.034 & 8,11,11, 9, 5, 6 \\
    17 & 611.51 & 506.30 &17.871&  1.063 & 0.250 & 2.572 &\nodata& 0.008&0.019&0.035&0.024&\nodata&14, 9, 5,11, 8, 0 \\
    18 & 611.53 & -52.85 &17.996&  0.596 & 0.218 & 2.678 & 0.877& 0.008&0.014&0.023&0.020&0.041 &14,10, 9,11, 8, 4 \\
    19 & 610.93 & 600.22 &17.634&  0.960 & 0.434 & 2.564 &\nodata& 0.009&0.022&0.034&0.015&\nodata&15,11, 8,12, 8, 0 \\
    20 & 610.80 & -78.72 &18.415&  0.615 & 0.114 & 2.636 & 0.600& 0.008&0.013&0.027&0.026&0.041 & 8,10, 9, 7, 3, 3 \\
    21 & 610.41 &-424.81 &18.194&  0.545 & 0.232 & 2.669 & 0.714& 0.017&0.025&0.032&0.019&0.038 & 7, 7, 8, 9, 6, 1 \\
    22 & 610.75 &-489.43 &16.983&  0.532 & 0.142 & 2.696 & 0.764& 0.005&0.008&0.013&0.007&0.017 &13,11,11,14,11, 8 \\
    23 & 610.28 & 402.95 &18.414&  0.588 & 0.146 & 2.677 & 0.695& 0.016&0.025&0.040&0.030&0.061 & 7,10, 7, 8, 5, 3 \\
    24 & 609.90 &-153.99 &17.819&  0.562 & 0.218 & 2.660 & 0.840& 0.007&0.011&0.019&0.020&0.032 &13,12,11,13, 7, 7 \\
    27 & 608.52 & 665.13 &18.128&  0.476 & 0.248 & 2.618 & 0.838& 0.011&0.015&0.021&0.026&0.025 &10, 9, 9,10, 6, 8 \\
\enddata
\end{deluxetable}

\newpage
\begin{deluxetable}{rrrrrrrrrrr}
\tablenum{3}
\tablecolumns{11}
\tablewidth{0pc}
\tablecaption{Photometric Index Averages for Turnoff Stars in NGC 6791}
\tablehead{
\colhead{$N$}     & 
\colhead{$V$}     & 
\colhead{$s.e.m.$}     & 
\colhead{$b-y$}     & 
\colhead{$s.e.m.$} &
\colhead{$m_1$} &
\colhead{$s.e.m.$}     & 
\colhead{H$\beta$}     & 
\colhead{$s.e.m.$}     & 
\colhead{$hk$}     & 
\colhead{$s.e.m.$}      }
\startdata
   23&17.349&0.0013 &0.589&0.0021& 0.214&0.0031& 2.580&0.0025 &0.843&0.0036 \\
   36&17.452&0.0012 &0.585&0.0018& 0.193&0.0026& 2.599&0.0021 &0.813&0.0029 \\
   40&17.546&0.0012 &0.573&0.0018& 0.197&0.0026& 2.591&0.0022 &0.805&0.0030 \\
   29&17.648&0.0014 &0.567&0.0021& 0.187&0.0031& 2.602&0.0027 &0.804&0.0036 \\
   25&17.748&0.0016 &0.567&0.0024& 0.188&0.0034& 2.595&0.0031 &0.795&0.0041 \\
   35&17.843&0.0014 &0.567&0.0021& 0.195&0.0030& 2.598&0.0026 &0.812&0.0036 \\
   31&17.944&0.0016 &0.569&0.0023& 0.189&0.0034& 2.594&0.0031 &0.798&0.0039 \\
   40&18.039&0.0016 &0.571&0.0022& 0.197&0.0032& 2.596&0.0030 &0.805&0.0039 \\
   34&18.149&0.0017 &0.584&0.0026& 0.201&0.0038& 2.607&0.0034 &0.818&0.0046 \\
   26&18.253&0.0021 &0.590&0.0030& 0.202&0.0044& 2.603&0.0043 &0.823&0.0056 \\
\enddata
\end{deluxetable}

\end{document}